\documentclass[10pt,letterpaper,twocolumn,ruled,vlined,linesnumbered,twocolumn]{IEEEtran}
\usepackage[utf8]{inputenc}
\usepackage{array}
\usepackage{verbatim}
\usepackage{multirow}
\usepackage{algorithm2e}
\usepackage{amsmath}
\usepackage{amssymb}
\usepackage{graphicx}

\makeatletter

\pdfpageheight\paperheight
\pdfpagewidth\paperwidth

\providecommand{\tabularnewline}{\\}

\usepackage{amsmath}
\usepackage{amsfonts}
\usepackage{amssymb}
\usepackage{amsthm}

\usepackage{graphicx}

\usepackage{multirow}
\usepackage{array}
\usepackage{slashbox}
\usepackage{makecell}
\usepackage{threeparttable}

\usepackage{rotating}

\usepackage{pict2e}
\usepackage{cases}
\usepackage{enumitem}
\usepackage{footnote}
\makesavenoteenv{tabular}
\makesavenoteenv{table}

\usepackage{floatrow}
\usepackage{color}
\usepackage{cite} 

\usepackage{url}

\newcommand{\figref}{Fig. }
\newcommand{\tabref}{Table }
\newcommand{\secref}{Section }
\newcommand{\algref}{Algorithm }

\newcommand{\define}[1]{{#1}}
\newcommand{\genmat}[1]{\mathbf{F}^{\otimes #1}}
\newcommand{\llr}[2]{L_{#2}^{#1}}
\newcommand{\pmc}[2]{\gamma _{#2}^{#1}}
\newcommand{\pmx}[2]{\gamma _{#2}^{'#1}}

\title{A High-Throughput Architecture of \\ List Successive Cancellation Polar Codes Decoder \\ with Large List Size}
\author{
ChenYang~Xia,~\IEEEmembership{Student Member,~IEEE,}
Ji~Chen,
YouZhe~Fan,~\IEEEmembership{Member,~IEEE,} \\
Chi-ying~Tsui,~\IEEEmembership{Senior Member,~IEEE,} 
Jie~Jin,
Hui~Shen,~\IEEEmembership{Member,~IEEE,}
 and Bin~Li,~\IEEEmembership{Member,~IEEE}
\thanks{C.-Y. Xia, J. Chen, Y.-Z. Fan, and C.-Y. Tsui 
are with the Department of Electronic and Computer Engineering, 
Hong Kong University of Science and Technology, Kowloon, Hong Kong 
(e-mail: cxia@connect.ust.hk; jchenbh@connect.ust.hk; jasonfan@connect.ust.hk; eetsui@ust.hk).}
\thanks{J. Jin, H. Shen, and B. Li are with the Communications Technology Research Laboratory, 
Huawei Technologies, Shenzhen, China 
(e-mail: steven.jinjie@huawei.com; henry.shenhui@huawei.com; binli.binli@huawei.com).}
}

\@ifundefined{showcaptionsetup}{}{%
 \PassOptionsToPackage{caption=false}{subfig}}
\usepackage{subfig}
\makeatother

\begin{document}
\maketitle

\begin{abstract}
As the first kind of forward error correction (FEC) codes that achieve
channel capacity, polar codes have attracted much research interest
recently. Compared with other popular FEC codes, polar codes decoded
by list successive cancellation decoding (LSCD) with a large list
size have better error correction performance. However, due to the
serial decoding nature of LSCD and the high complexity of list management
(LM), the decoding latency is high, which limits the usage of polar
codes in practical applications that require low latency and high
throughput. In this work, we study the high-throughput implementation
of LSCD with a large list size. Specifically, at the algorithmic level,
to achieve a low decoding latency with moderate hardware complexity,
two decoding schemes, a multi-bit double thresholding scheme and a
partial G-node look-ahead scheme, are proposed. Then, a high-throughput
VLSI architecture implementing the proposed algorithms is developed
with optimizations on different computation modules. From the implementation
results on UMC 90 nm CMOS technology, the proposed architecture achieves
decoding throughputs of 1.103 Gbps, 977 Mbps and 827 Mbps when the
list sizes are 8, 16 and 32, respectively.
\end{abstract}


\section{Introduction\label{sec:introduction}}

\IEEEPARstart{P}{olar} codes are the first kind of \define{forward error correction}
(FEC) codes that provably achieve channel capacity \cite{earikan_bilkent_tit_2009_polar,ital_ucsd_tit_2013_construction}.
Their regular and low-complexity decoding algorithm is hardware-friendly
and hence polar codes have attracted much research interest recently. 

\define{Successive cancellation decoding} (SCD) \cite{amishra_epfl_asscc_2012_asic,cleroux_mcgill_tsp_2013_semiparallel,czhang_umn_tcasii_2014_simplified,yzfan_hkust_tsp_2014_effps,czhang_umn_icc_2012_lookahead,czhang_umn_tsp_2013_overlap,byuan_umn_tcasi_2014_sc2bd,gsarkis_mcgill_jsac_2014_fast}
and \define{belief propagation decoding} (BPD) \cite{apamuk_icwcs_bilkent_2011_fpga,byuan_umn_tsp_2014_earlystop,smabbas_hkust_tvlsi_2016_bpd}
are the two main kinds of decoding schemes for polar codes. The complexity
of SCD is low but the decoding is sequential in nature, and therefore
it is a challenge to achieve a low latency and a high throughput.
Recently, research work has been done to improve the latency of SCD
\cite{czhang_umn_icc_2012_lookahead,czhang_umn_tsp_2013_overlap,byuan_umn_tcasi_2014_sc2bd,gsarkis_mcgill_jsac_2014_fast}.
BPD uses message passing among the nodes of the factor graph of the
polar codes to carry out decoding. It is parallel in nature and consequently
has a high decoding throughput. However, its hardware cost is large
and more importantly, its error correction performance is not as good
as that of SCD. In this work, we mainly focus on SCD-based decoding
schemes of polar codes.%

Comparing the error correction performance of SCD on polar codes with
that of the start-of-the-art FEC codes, such as \define{low-density parity-check}
(LDPC) codes \cite{rggallager_mit_book_1963_ldpc} or turbo codes
\cite{cberrou_bretagne_icc_1993_turbo}, polar codes with a short
code length are inferior. One method to improve the error correction
performance is to use a long code length because the channel polarization
phenomenon increases the ratio of almost-lossless channels for long-length
codes \cite{earikan_bilkent_tit_2009_polar,cleroux_mcgill_tsp_2013_semiparallel,gsarkis_mcgill_jsac_2014_fast}.
However, the decoding latency, which is related to the code length,
is high and so are the computation and memory overhead \cite{cleroux_mcgill_tsp_2013_semiparallel,yzfan_hkust_tsp_2014_effps}. 

Another method is using \define{list successive cancellation decoding}
(LSCD) \cite{kchen_bupt_iet_2012_lscd,ital_ucsd_tit_2015_list} in
which $\mathcal{L}$ SCDs are executed in parallel for decoding one
codeword. The $\mathcal{L}$ best decoding paths are kept during the
decoding and finally the path that satisfies the \define{cyclic redundancy check}
(CRC) \cite{ital_ucsd_tit_2015_list,bli_huawei_cl_2012_crc,kniu_bupt_cl_2012_crc}
is selected as the output. From the results shown in \cite{ital_ucsd_tit_2015_list,kniu_bupt_icc_2013_beyondturbo,bli_huawei_cl_2012_crc,kniu_bupt_cl_2012_crc},
with a \define{large list size} ($\mathcal{L}\geq16$), CRC-concatenated
polar codes using LSCD out-perform other state-of-the-art FEC codes.
However, LSCD with a large list size has significant latency and complexity
overhead on hardware. Thus, special algorithmic and architectural
optimization techniques are required to reduce the latency and complexity,
particularly for latency-sensitive applications such as the next generation
communication systems \cite{3gpp_3gpp_ran087_2016_5g}.

The first hardware architecture of LSCD was proposed in \cite{abalatsoukas_epfl_tcasii_2014_archlscd}
in which \define{log-likelihood} (LL) values are used for computing
the decoding messages. In \cite{byuan_umn_asilomar_2014_llrlscd,abalatsoukas_epfl_icassp_2014_llrlscd,abalatsoukas_epfl_tsp_2015_llrlscd,pgiard_epfl_jetcas_2017_polarbear},
\define{log-likelihood ratios} (LLRs) are used instead of LLs to
reduce the computational circuit complexity and memory usage. Most
of the LSCD architectures have two main processing modules. The first
is the \define{list management} (LM) module, which maintains the
list when the expanded list size exceeds $\mathcal{L}$; and the second
module consists of $\mathcal{L}$ parallel SCD cores that calculate
the messages of each path simultaneously. To achieve a low-latency
and high-throughput design, both modules need to be optimized.

During the decoding of every bit, $\mathcal{L}$ survival paths are
expanded to $2\mathcal{L}$ paths and the LM module is responsible
for selecting the best $\mathcal{L}$ paths to keep. To reduce the
decoding latency, two classes of decoding schemes originally proposed
for single SCD \cite{byuan_umn_tcasi_2014_sc2bd,gsarkis_mcgill_jsac_2014_fast},
which decode multiple bits at the same time, are adopted for LSCD.
\begin{itemize}
\item The first is \define{multi-bit decoding} (MBD) \cite{byuan_umn_tvlsi_2015_sclmbd,byuan_umn_tcasii_2017_sclmbd,crxiong_lehigh_sips_2014_symbol,crxiong_lehigh_tsp_2016_symbol},
extended from \cite{byuan_umn_tcasi_2014_sc2bd}. Here, the number
of bits that are decoded together in each MBD, $M$, is a fixed value.
To simplify the complexity of the sorter by which at most $2^{M}\cdot\mathcal{L}$
numbers need to be sorted, a two-stage sorting strategy was proposed
in \cite{crxiong_lehigh_tsp_2016_symbol}, where a local sorting is
done in each path at the first stage and only a few best local paths
are sent to the second stage for global sorting and selection.
\item The second class of decoding schemes \cite{sahashemi_mcgill_isit_2016_sscl,sahashemi_mcgill_tcasi_2016_ssclspc,sahashemi_mcgill_wcncw_2017_fastsscl,sahashemi_mcgill_tsp_2017_fastflexible,jlin_lehigh_sips_2014_rdclat,jlin_lehigh_tvlsi_2016_highthpt,gsarkis_mcgill_jsac_2016_sclfast}
is based on the concept of \define{fast simplified successive cancellation}
(fast-SSC) decoding \cite{gsarkis_mcgill_jsac_2014_fast}. For simplicity,
we call these fast-SSC-based algorithms. Here, a block of code is
divided into several kinds of special sub-codes that can be decoded
by simplified decoding algorithms. Unlike MBD, the number of bits
that can be decoded together is not fixed. In each step, at most $2\mathcal{L}$
paths are expanded, and consequently the complexity of the sorter
is restricted.
\end{itemize}
The complexity of the sorter also increases dramatically with $\mathcal{L}$
and hence will incur a large delay overhead when $\mathcal{L}$ is
large. In \cite{abalatsoukas_epfl_tcasii_2014_archlscd} and \cite{abalatsoukas_epfl_tsp_2015_llrlscd},
a parallel radix-$2\mathcal{L}$ sorter is used to sort and select
the best $\mathcal{L}$ paths to reduce the logic delay. However,
from the result in \cite{abalatsoukas_epfl_iscas_2015_sorting,sahashemi_mcgill_tsp_2017_fastflexible},
when $\mathcal{L}\geq8$, the logic delay of the sorter becomes critical
and dictates the clock rate. Consequently, most of the implementation
of the existing ASIC architectures only present results for LSCD with
$\mathcal{L}\leq8$ \cite{jlin_lehigh_sips_2014_rdclat,abalatsoukas_epfl_tsp_2015_llrlscd,byuan_umn_tvlsi_2015_sclmbd,byuan_umn_tcasii_2017_sclmbd,jlin_lehigh_tvlsi_2016_highthpt,crxiong_lehigh_sips_2014_symbol,crxiong_lehigh_tsp_2016_symbol,sahashemi_mcgill_tsp_2017_fastflexible,sahashemi_mcgill_tcasi_2016_ssclspc,pgiard_epfl_jetcas_2017_polarbear,sahashemi_mcgill_wcncw_2017_fastsscl,sahashemi_mcgill_isit_2016_sscl},
and these architectures are not suitable for LSCD with a large list
size. In our previous work \cite{yzfan_hkust_icassp_2015_dts,yzfan_hkust_jsac_2016_sedts},
a \define{double thresholding scheme} (DTS) was proposed, in which
an approximate sorting method is used with the help of two run-time
generated threshold values. By doing this, the sorting and hence the
selection of the best $\mathcal{L}$ paths do not scale with $\mathcal{L}$
and the scheme is suitable for LSCD with a large list size. Implementation
results show that the architecture of LSCD with $\mathcal{L}=16$
using the DTS doubled the decoding throughput and the list size when
compared with the state-of-the-art architectures at that time. For
LSCD with $\mathcal{L}=32$, only CPU-based \cite{gsarkis_mcgill_jsac_2016_sclfast}
and FPGA-based \cite{cxia_hkust_fpl_2017_fpgalarge} architectures
have been proposed and no ASIC architecture has been reported. Further
optimization methods that selectively expand the paths at each bit
and eliminate unnecessary execution times of LM operations \cite{jlin_lehigh_tvlsi_2016_highthpt,yzfan_hkust_jsac_2016_sedts,cxia_hkust_fpl_2017_fpgalarge,zyzhang_zju_jsac_2016_splitreduce}
were proposed to reduce the overall latency. 

For the SCD cores, optimization methods used for traditional SCD are
still applicable to the LSCD architecture. A hardware multiplexing
scheme was first proposed in \cite{czhang_umn_tsp_2013_overlap} for
SCD and later was adapted to LSCD in \cite{tbchen_tamu_iscas_2016_scloverlap}.
With deep-pipelining, the LSCD can decode different paths of a frame
in sequential instead of in parallel. Theoretically, only one SCD
core is needed and the hardware complexity is only about $1/\mathcal{L}$
of that of the traditional LSCD. However, the latency is very long
and is not suitable for latency-sensitive applications. Another family
of ideas is pre-computation. In \cite{czhang_umn_icc_2012_lookahead},
a \define{G-node look-ahead} (GLAH) schedule was used, which pre-computed
the intermediate LLRs in SCD to save the latency in the decoding process,
and the corresponding register-based architecture was proposed. In
\cite{crxiong_lehigh_tsp_2016_symbol}, a \define{pre-computation memory-saving}
(PCMS) scheme was proposed to save the memory for storing the channel
LLRs. 

In this work, we focus on the high-throughput design of LSCD with
a large list size due to its extraordinary error correction performance.
Two algorithmic level techniques are proposed to reduce the decoding
latency of LSCD and the corresponding hardware architecture is designed,
of which the critical path delay is optimized. Specifically, our main
contributions are as follows. 
\begin{itemize}
\item To reduce the latency of the LM module, a method called \define{multi-bit double thresholding scheme}
(MB-DTS) is proposed, which combines the idea of MBD with DTS and
\define{selective expansion} \cite{yzfan_hkust_jsac_2016_sedts}.
The original DTS is executed bit by bit and the whole scheduling tree
needs to be traversed, which results in a large decoding latency.
The proposed method executes DTS for all the bits in a sub-tree simultaneously
to avoid the full tree traversal. Thus, a much lower latency is achieved
while the low-complexity characteristic of DTS is maintained. 
\item For the SCD module, the idea of \define{partial G-node look-ahead}
(P-GLAH) is proposed and adopted in a semi-parallel architecture \cite{abalatsoukas_epfl_tsp_2015_llrlscd,yzfan_hkust_jsac_2016_sedts}.
The original GLAH for a single SCD can save half of the clock cycles
for decoding a block of code. However, it uses three times as much
memory as the conventional semi-parallel architecture. In this work,
we make a careful tradeoff between decoding latency and the hardware
usage of GLAH, and the proposed P-GLAH technique has a similar latency
as GLAH while the hardware overhead is minimized when comparing with
the counterpart in the traditional LSCD architecture. 
\item A high-throughput architecture that is suitable for LSCD with a large
list size is proposed based on the proposed algorithms. The structures
of the programmable processing element and the path metric update
block used in the architecture are carefully designed for a short
logic delay. A critical path delay optimization scheme is proposed
to further reduce the decoding latency. Experimental results show
that high-throughput LSCD is achievable even with a list size of 32.
\end{itemize}

The rest of this paper is organized as follows. In \secref\ref{sec:preliminaries},
the background and the existing decoding methods of polar codes will
be reviewed. In \secref\ref{sec:mbdts}, the algorithm and the detailed
decoding schedule of the proposed MB-DTS will be discussed. In \secref\ref{sec:pglah},
the P-GLAH technique for the SCD cores will be presented. In \secref\ref{sec:arch_llls},
the high-throughput architecture for LSCD with a large list size will
be introduced. Finally, the experimental results of the implementation
of LSCDs with large list sizes will be presented in \secref\ref{sec:experiment},
and conclusions will be given in \secref\ref{sec:conclusion}.

\section{Preliminaries\label{sec:preliminaries}}

\subsection{Construction of Polar Codes\label{subsec:polar}}

Polar codes are a family of block codes\cite{earikan_bilkent_tit_2009_polar}.
Let $N=2^{n}$ be the code length of polar codes, $\mathbf{u}_{N}$
and $\mathbf{x}_{N}$ (each is an $N$-bit vector) be the input source
word and the output codeword, respectively, and the encoding of polar
codes is given by
\begin{equation}
\mathbf{x}_{N}=\mathbf{u}_{N}\cdot\genmat{n},\label{eq:polar_gen}
\end{equation}
where $\genmat{n}$ is the $n^{th}$ Kronecker power of $\mathbf{{F}=\begin{bmatrix}\begin{array}{cc}
1 & 0\\
1 & 1
\end{array}\end{bmatrix}}$. Due to the polarization effect, some of the $N$ bits are more reliable
and hence are used to transmit information and are called the information
bits, while the other not-so-reliable bits are set to 0 and are called
the frozen bits. $\mathcal{\mathcal{A}}$ and $\mathcal{A}^{c}$ denote
the sets of all the indices of the information and frozen bits, respectively.
$R=K/N$ is defined as the code rate of polar codes, where $K$ is
the cardinality of $\mathcal{\mathcal{A}}$. If an $r$-bit CRC code
is used, the last $r$ information bits are used to transmit the checksum
generated from the other $K-r$ information bits\footnote{It is noted that the $r$ parity check bits do not truly transmit
information and the code rate is correspondingly modified to $(K-r)/N$.
However, as they are treated in the same way as information bits from
the perspective of polar code decoder, we still regard them as information
bits in this paper.}. 

\subsection{Successive Cancellation Decoding\label{subsec:scd}}

\begin{figure}
\includegraphics[width=8.8cm]{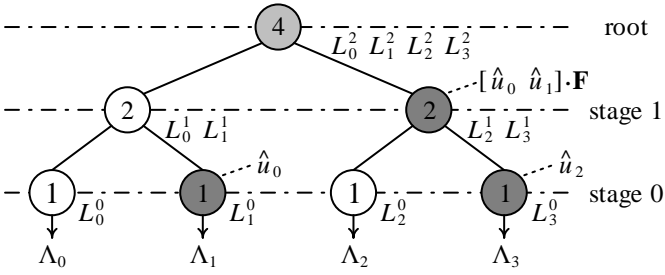}\caption{Scheduling tree of SCD for polar codes with $N=4$.}
\label{fig:sch_tree_4}
\end{figure}

SCD is the most popular decoding method of polar codes due to its
low complexity which is $\mathcal{O}(N\log N)$. The decoding process
can be represented by a scheduling tree. The scheduling tree for $N=4$
polar codes is shown in \figref \ref{fig:sch_tree_4} as an example.
It has $n+1$ stages with descending indices from the root to the
leaves. The number of nodes at stage $s$ of the scheduling tree is
$2^{n-s}$, while the number of functions executed in each node is
$2^{s}$, where $s\in[0,n]$. In each stage, there are $N$ output
LLR operands. We hereby denote the $i^{th}$ LLR at stage $s$ as
$\llr{s}{i}$, where $i\in[0,N-1]$. 

At the root node of the scheduling tree, $N$ channel LLRs are inputted
to the tree and are expressed as
\begin{equation}
\llr{n}{i}=\log(\text{Pr}(\mathbf{y}|0))-\log(\text{Pr}(\mathbf{y}|1)),i\in[0,N-1],\label{eq:llr}
\end{equation}
where $\textbf{y}$ is the channel outputs on $\textbf{x}$. At the
lowest stage, $\Lambda_{i}=\llr{0}{i}$ is the LLR corresponding to
bit $u_{i}$. For $i\in\mathcal{A}^{c}$, the decoded value of $u_{i}$,
denoted by $\hat{u}_{i}$, is always equal to 0; while for $i\in\mathcal{A}$,
$\hat{u}_{i}$ is evaluated according to
\begin{equation}
\hat{u}_{i}=\Theta(\Lambda_{i})=\begin{cases}
0, & \textrm{if}\,\Lambda_{i}>0,\\
1, & \text{otherwise.}
\end{cases}\label{eq:sgn_func}
\end{equation}
To calculate $\Lambda_{i}$s, LLR calculations are executed in each
node of the scheduling tree. Specifically, there are two kinds of
nodes: F-nodes and G-nodes, which are represented by the white and
black circles in \figref \ref{fig:sch_tree_4}, respectively. During
the decoding process, the following calculations are executed in the
F- and G-nodes, respectively:
\begin{align}
L_{\text{F}}(L_{a},L_{b}) & =2\text{tanh}^{-1}(\text{tanh}(L_{a}/2)\cdot\text{tanh}(L_{b}/2)),\label{eq:f_func}\\
L_{\text{G}}(L_{a},L_{b},\hat{ps}) & =(-1)^{\hat{ps}}L_{a}+L_{b},\label{eq:g_func}
\end{align}
where $L_{a}$ and $L_{b}$ are the two input LLRs to the node and
$\hat{ps}$ is the \define{partial-sum}. The partial-sums at stage
$s$ are obtained by
\begin{equation}
[\hat{ps}_{(2j+1)\cdot2^{s}}^{s},..,\hat{ps}_{(2j+2)\cdot2^{s}-1}^{s}]=[\hat{u}_{2j\cdot2^{s}},..,\hat{u}_{(2j+1)\cdot2^{s}-1}]\cdot\genmat{s},\label{eq:ps}
\end{equation}
where $j\in[0,2^{n-s-1}]$. It can be seen that the input $\hat{ps}$
value for a particular G-node depends on the values of the bits that
are already decoded. Therefore, the G-node cannot be computed until
all the corresponding leaf nodes have been visited. Thus, the decoding
process of SCD can be represented by a depth-first traversal of the
scheduling tree.

To simplify the hardware implementation, usually \eqref{eq:f_func}
is approximated using a \define{min-sum} calculation which is given
by
\begin{equation}
L_{\text{F,ms}}(L_{a},L_{b})\approx(-1)^{\Theta(L_{a})\oplus\Theta(L_{b})}\cdot\text{min}(|L_{a}|,|L_{b}|).\label{eq:min-sum}
\end{equation}

\subsection{List Successive Cancellation Decoding\label{subsec:lscd}}

To improve the error correction performance of SCD, list successive
cancellation decoding was proposed \cite{ital_ucsd_tit_2015_list,kchen_bupt_iet_2012_lscd}.
LSCD keeps $\mathcal{L}$ paths of decoded bits during decoding, where
each path is denoted as $\hat{\mathbf{u}}_{i}^{l}$, $l\in[0,\mathcal{L}-1]$
and $i$ is the index of the bit just decoded. To decode $\mathcal{L}$
paths in parallel, $\mathcal{L}$ copies of SCD are used. 

At the beginning of decoding, only one path is valid. When the decoding
process reaches a leaf node corresponding to an information bit $u_{i}$
$(i\in\mathcal{A})$, a path is expanded to two by keeping both possible
values of the bit, and the number of valid paths in the list doubles.
This number increases exponentially with respect to the number of
decoded information bits, and the list becomes full after $\log\mathcal{L}$
information bits are decoded. In the subsequent decoding, an LM operation
is required to keep the list size to $\mathcal{L}$ based on the \define{path metrics}
(PMs) of all the paths. Let $\hat{\mathbf{u}}_{i+1}^{'k}$\footnote{The apostrophe here means the path is an expanded one.}
be the the $k^{th}$ path expanded from the $l^{th}$ path $\hat{\mathbf{u}}_{i}^{l}$
which is resulted from the decoding of the $i^{th}$ $(i\in\mathcal{A})$
bit. Then, it is expressed as
\begin{equation}
\hat{\textbf{u}}_{i+1}^{'k}=[\hat{\textbf{u}}_{i}^{l},\hat{u}_{i}^{'k}]=[\hat{u}_{0}^{l},\hat{u}_{1}^{l},...,\hat{u}_{i-1}^{l},\hat{u}_{i}^{'k}],\label{eq:path_ext}
\end{equation}
where $k=2l\text{ or }2l+1$ as it is extended from path $l$ and
$\hat{u}_{i}^{'k}$ is the value of the latest decoded bit (either
0 or 1). The PM of an expanded path is updated based on bit-wise accumulation
and is given by
\begin{equation}
\pmx{k}{i+1}=\pmc{l}{i}+\log\{1+\exp[(2\hat{u}_{i}^{'k}-1)\cdot\Lambda_{i}^{l}]\},\label{eq:pm_update}
\end{equation}
where $\pmc{l}{i}$ and $\pmx{k}{i+1}$ are the PMs of the original
path $\hat{\mathbf{u}}_{i}^{l}$ and the expanded path $\hat{\mathbf{u}}_{i+1}^{'k}$,
respectively; and $\Lambda_{i}^{l}$ is the LLR output $\llr{0}{i}$
at the leaf node $u_{i}$ of the $l^{th}$ original path. For easier
hardware implementation, the \define{path metric update} (PMU) operation
in \eqref{eq:pm_update} is approximated as
\begin{equation}
\begin{cases}
\pmx{2l}{i+1}=\pmc{l}{i}, & \textrm{where}~\hat{u}_{i}^{'2l}=\Theta\left(\Lambda_{i}^{l}\right),\\
\pmx{2l+1}{i+1}=\pmc{l}{i}+|\Lambda_{i}^{l}|, & \textrm{where}~\hat{u}_{i}^{'2l+1}=1-\Theta\left(\Lambda_{i}^{l}\right),
\end{cases}\label{eq:pmu_aprx}
\end{equation}
where the sign function $\Theta(x)$ returns the sign bit of $x$.
So if a newly decoded bit has the same sign as $\Lambda_{i}^{l}$,
the PM value of the expanded path will be the same as that of the
original path and $k$ is set as $2l$; otherwise, a \define{penalty}
of $|\Lambda_{i}^{l}|$ will be added to the current PM value and
$k$ is set as $2l+1$. For $i\in\mathcal{A}$, both of the expressions
in \eqref{eq:pmu_aprx} will be computed. After that, list pruning
will be executed, where the PMs of the $2\mathcal{L}$ expanded paths
will be sorted and the $\mathcal{L}$ paths with the smaller PMs will
be kept in the list. For $i\in\mathcal{A}^{c}$, $\hat{u}_{i}\equiv0$,
hence only one of \eqref{eq:pmu_aprx} will be computed and list pruning
is not needed as the list size is still equal to $\mathcal{L}$.
\begin{figure}
\includegraphics[width=8.4cm]{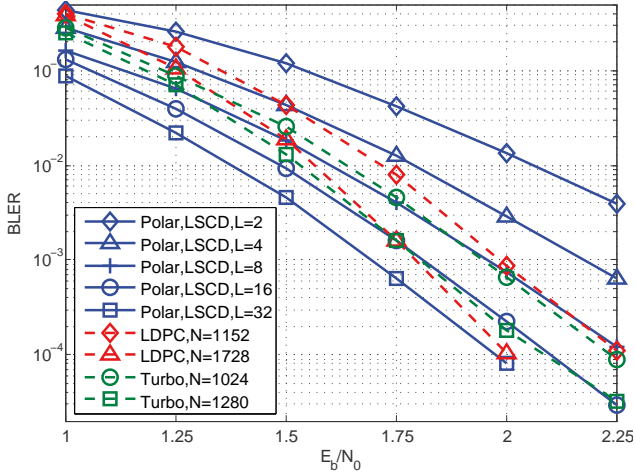}

\caption{Comparison of error correction performance of polar codes, LDPC codes
(decoded by modified belief propagation with 50 iterations) and turbo
codes (decoded by log maximum a posteriori with 8 iterations).}
\label{fig:err_lscd_ldpc}
\end{figure}

Usually, LSCD with a larger list size has a better error correction
performance. Figure \ref{fig:err_lscd_ldpc} shows the simulation
results of LSCDs with different list sizes for the CRC-concatenated
polar code of $(N,K,r)=(1024,512,24)$ under \define{additive white Gaussian noise}
(AWGN) channel\footnote{CRC-24-Radix-64 with generator polynomial equals to 0x1864cfb}.
As reference, the error correction performances of rate-$\frac{1}{2}$
LDPC codes in IEEE 802.16e standard \cite{mhjang_snu_isita_2006_ldpcpfm}
and turbo codes in LTE standard \cite{kniu_bupt_icc_2013_beyondturbo}
with different code lengths are also presented. It can be seen that
the \define{block error rates} (BLERs) of LSCDs with $\mathcal{L}=16$
and 32 out-perform those of LDPC codes with similar code lengths.
However, as discussed in \secref \ref{sec:introduction}, the computational
load of LM and hence the decoding latency of LSCD are greatly increased
when a large list size is used. In the next sub-section, we review
some existing low-latency decoding algorithms for LSCD.%

\subsection{Review of Some of the Existing Low-latency Decoding Algorithms for
LSCD\label{subsec:review-lowlat}}

\subsubsection{Double Thresholding Scheme \label{subsub:review-dts}}

First, we briefly review the basic idea of DTS proposed in \cite{yzfan_hkust_jsac_2016_sedts}.
As mentioned in \secref \ref{subsec:lscd}, the number of expanded
paths is $2\mathcal{L}$ after an information bit is decoded. For
list pruning, a $2\mathcal{L}$-to-$\mathcal{L}$ sorter is needed
to sort all the PMs and selects the $\mathcal{L}$ paths with the
smallest PMs. The hardware complexity of a parallel sorter is $\mathcal{O}(\mathcal{L}^{2})$
and becomes very large when a large list size $\mathcal{L}$ is used.
The critical path delay is also increased due to the high complexity
and thus influences the throughput of the decoder.

To remove the $2\mathcal{L}$-to-$\mathcal{L}$ sorter, DTS was proposed
for list pruning. Suppose that the $\mathcal{L}$ PM values of the
original paths after decoding $\hat{u}_{i-1}$, $[\pmc{0}{i},\pmc{1}{i},...,\pmc{\mathcal{L}-1}{i}]$,
are already sorted, then two threshold values, an acceptance threshold
$AT$ and a rejection threshold $RT$, can be obtained as follows:
\begin{equation}
[AT,RT]=[\pmc{\mathcal{L}/2}{i},\pmc{\mathcal{L}-1}{i}].\label{eq:atrt}
\end{equation}
It is proved in \cite{yzfan_hkust_jsac_2016_sedts} that any path
that has a PM value smaller than $AT$ is a survival path after pruning
and no survival path has a PM value larger than $RT$. Using these
two thresholds, $\mathcal{L}$ $\pmc{l}{i+1}$s are selected from
$2\mathcal{L}$ $\pmx{k}{i+1}$s based on the following criteria without
the need of sorting the expanded PMs:
\begin{itemize}
\item If $\pmx{k}{i+1}<AT$, the corresponding path is kept;
\item If $\pmx{k}{i+1}>RT$, the corresponding path is pruned;
\item For the paths with $AT<\pmx{k}{i+1}<RT$ , they are randomly chosen
such that the list is filled with $\mathcal{L}$ paths.
\end{itemize}
From the simulation results in \cite{yzfan_hkust_jsac_2016_sedts},
the DTS has negligible performance degradation, while only $\mathcal{O}(\mathcal{L})$
comparisons with two threshold values are needed to select the path
without using a sorting operation. This results in low logic delay
and allows the DTS to be finished in one cycle. It is noted that to
extract the two thresholds for the next DTS, the survival PMs according
to the above criteria need to be sorted. However, this sorting is
performed on fewer PMs than that in traditional LM, so it can be executed
in parallel with the SCD core computation\cite{yzfan_hkust_jsac_2016_sedts}
and finished before the next DTS operation. Thus the decoding latency
will not be affected.

\subsubsection{Selective Expansion\label{subsub:review-se}}

Suppose all the bits in a codeword are sorted according to their reliability
with respect to the \define{bit error rate} (BER). The construction
of the polar codes uses the $K$ bits with high reliability as the
information bits, and the rest as the frozen bits. However, the reliability
of these $K$ information bits is not the same and the reliability
of some information bits is higher than the others. Based on this
property, a low-latency LM scheme called selective expansion is proposed
in \cite{yzfan_hkust_jsac_2016_sedts}. With selective expansion,
the information set $\mathcal{A}$ is further divided into two sub-sets
– a reliable set, denoted as $\mathcal{A}_{r}$, and an unreliable
set, denoted as $\mathcal{A}_{u}$. The bits in $\mathcal{A}_{u}$
are those with relatively low reliability among all of the information
bits. The path expansion and PMU for these bits are the same as those
described in \eqref{eq:path_ext} and \eqref{eq:pmu_aprx}. However,
for the bits in $\mathcal{A}_{r}$ which have higher reliability,
we only keep the path with the extended bit that matches with the
hard decision result of the decoding, and hence the path expansion
is simplified as
\begin{equation}
\hat{\textbf{u}}_{i+1}^{l}=[\hat{u}_{0}^{l},...,\hat{u}_{i}^{l}]=[\hat{u}_{0}^{l},...,\hat{u}_{i-1}^{l},\Theta(\Lambda_{i}^{l})],\label{eq:path_ext_ar}
\end{equation}
and the PMs are not updated, i.e.,
\begin{equation}
\pmc{l}{i+1}=\pmc{l}{i}.\label{eq:pmu_ar}
\end{equation}
Effectively, path expansion and PMU can be omitted. Thus, the complexity
of decoding the reliable bits in LSCD is similar to that of the frozen
bits, as summarized in \tabref \ref{tab:se}. It is demonstrated
in \cite{yzfan_hkust_jsac_2016_sedts} that a carefully chosen $\mathcal{A}_{r}$
can bring a great reduction in the execution times of LM operations,
while the error correction performance degradation is negligible.
\begin{table}
\caption{Three kinds of bits in an LSCD with selective expansion}
\label{tab:se}

{\small{}}%
\begin{tabular}{c|c|c|c}
\hline 
 & \multicolumn{3}{c}{{\small{}$\leftarrow$ Low ............ BER ............ High $\rightarrow$}}\tabularnewline
\hline 
\multirow{2}{*}{{\small{}Set}} & \multicolumn{2}{c|}{{\small{}Information $\mathcal{A}$}} & \multirow{2}{*}{{\small{}Frozen $\mathcal{A}^{c}$}}\tabularnewline
\cline{2-3} 
 & {\small{}Reliable $\mathcal{A}_{r}$} & {\small{}Unreliable $\mathcal{A}_{u}$} & \tabularnewline
\hline 
{\small{}Path expansion} &  & {\small{}$\checkmark$} & \tabularnewline
\hline 
{\small{}PMU} &  & {\small{}$\checkmark$} & {\small{}$\checkmark$}\tabularnewline
\hline 
{\small{}Decoded bits} & {\small{}0 or 1} & {\small{}0 and 1} & {\small{}0}\tabularnewline
\hline 
\end{tabular}{\small \par}
\end{table}

\subsubsection{Multi-bit Decoding and Fast-SSC-based Methods\label{subsub:review-mbd}}

To reduce the decoding latency, multiple bits in a block of polar
code can be decoded at the same time. MBD and fast-SSC-based methods
are the two most popular classes of decoding algorithms in the literature
that are based on this idea. 

In MBD \cite{byuan_umn_tvlsi_2015_sclmbd,byuan_umn_tcasii_2017_sclmbd,crxiong_lehigh_sips_2014_symbol,crxiong_lehigh_tsp_2016_symbol},
suppose that $M$ bits are decoded simultaneously and hence the list
is only updated once during this process, the corresponding leaf nodes
of the $M$ bits in the scheduling tree have the same root node at
stage $m$, where $m=\log_{2}M$. Considering the worst case where
all the $M$ bits are information bits, $2^{M}$ paths are expanded
from any of the $\mathcal{L}$ existing paths and the PM of one of
these $2^{M}$ paths is updated as
\begin{equation}
\pmx{l\cdot2^{M}+V}{i+M}=\pmc{l}{i}+\sum_{j=0}^{M-1}(v_{j}\oplus\Theta((\llr{m}{j})^{l}))\cdot|(\llr{m}{j})^{l}|,\label{eq:mbd_pmu}
\end{equation}
where $\{(\llr{m}{0})^{l},(\llr{m}{1})^{l},...,(\llr{m}{M-1})^{l}\}$
are the LLR inputs at the root of the corresponding sub-tree at stage
$m$ of the $l^{th}$ path, $v_{0}v_{1}...v_{M-1}$ is the binary
vector representation of $V\in[0,2^{M}-1]$, which is actually encoded
from the $M$ evaluated bits rooted at this sub-tree and is given
by
\begin{equation}
[v_{0},...,v_{M-1}]=[\hat{u}_{i}^{'l\cdot2^{M}+V},...,\hat{u}_{i+M-1}^{'l\cdot2^{M}+V}]\cdot\genmat{m},\label{eq:mbd_evl}
\end{equation}
and $[\hat{u}_{i}^{'l\cdot2^{M}+V},...,\hat{u}_{i+M-1}^{'l\cdot2^{M}+V}]$
is a sub-vector of the decoded bits of the expanded path, which is
given by
\begin{equation}
\hat{\textbf{u}}_{i+M}^{'l\cdot2^{M}+V}=[\hat{u}_{0}^{l},...,\hat{u}_{i-1}^{l},\hat{u}_{i}^{'l\cdot2^{M}+V},...,\hat{u}_{i+M-1}^{'l\cdot2^{M}+V}].\label{eq:mbd_path}
\end{equation}
Since we have $\mathcal{L}$ original paths, $\mathcal{L}\cdot2^{M}$
expanded paths are generated, and \eqref{eq:mbd_pmu} is executed
$\mathcal{L}\cdot2^{M}$ times correspondingly. To keep the list size
as $\mathcal{L}$, list pruning is executed in the same way as that
of the traditional LSCD. 

If an $M$-bit pattern contains $M_{\text{frz}}$ frozen bits, the
number of expanded paths will be reduced to $2^{M_{\text{inf}}}$,
where $M_{\text{inf}}=M-M_{\text{frz}}$. For such a pattern, the
computational load of \eqref{eq:mbd_pmu}, \eqref{eq:mbd_evl} and
\eqref{eq:mbd_path} can be reduced. However, the hardware has to
cater for the the worst case and hence an $\mathcal{L}\cdot2^{M}$-to-$\mathcal{L}$
sorter is required to execute list pruning, which leads to a much
higher hardware complexity than traditional LM. This is the main drawback
of MBD for LSCD.

Different from MBD, fast-SSC-based methods \cite{sahashemi_mcgill_isit_2016_sscl,sahashemi_mcgill_tcasi_2016_ssclspc,sahashemi_mcgill_wcncw_2017_fastsscl,sahashemi_mcgill_tsp_2017_fastflexible,jlin_lehigh_sips_2014_rdclat,jlin_lehigh_tvlsi_2016_highthpt,gsarkis_mcgill_jsac_2016_sclfast}
divide a block of polar code into four kinds of special sub-codes:
rate-0 code, repetition code (the last bit is an information bit and
all the others are frozen bits), \define{single parity check} (SPC)
code (the first bit is a frozen bit and all the others are information
bits), and rate-1 code, with variable length. A simplified LM algorithm
for each kind of sub-code was proposed. For rate-0 code and repetition
code, only one or two paths are expanded from each survival path,
so the best $\mathcal{L}$ paths can still be picked by a $2\mathcal{L}$-input
sorter. The corresponding PMU algorithms are given in \cite{sahashemi_mcgill_isit_2016_sscl,sahashemi_mcgill_tcasi_2016_ssclspc}.
For SPC code and rate-1 code, mathematical proof has been given in
\cite{sahashemi_mcgill_wcncw_2017_fastsscl,sahashemi_mcgill_tsp_2017_fastflexible}
that the number of path expansions for which no performance degradation
is incurred is $\text{min}(M,\mathcal{L}-1)$, which means fewer path
expansions are needed when the list size is smaller than the actual
size of an SPC or a rate-1 code. 

\section{Multi-bit Double Thresholding Scheme \label{sec:mbdts}}

To reduce the decoding latency of LSCD with a large list size, we
propose to combine double thresholding scheme, selective expansion
and multi-bit decoding together. However, the original DTS requires
bit-wise expansions and each leaf node in the scheduling tree has
to be traversed. To apply MBD, special consideration has to be made.
We first show a method which uses DTS for a multi-bit tuple instead
of a single bit. Then we will present a low-latency LSCD algorithm
called multi-bit double thresholding scheme. The error correction
performance and complexity will then be discussed and compared with
the traditional DTS.

\subsection{Decoding Multi-bit Pattern with Single Unreliable Bit \label{subsec:subt}}

The original DTS selects the $\mathcal{L}$ best paths out of the
$2\mathcal{L}$ expanded paths when one information bit is expanded.
With selective expansion, we only need to consider the unreliable
bits. If we want to use DTS for decoding two unreliable bits, the
number of expanded paths will be $4\mathcal{L}$. In this case, $\pmc{\mathcal{L}/4}{i}$
will be used as $AT$\footnote{According to Proposition 1 in \cite{yzfan_hkust_jsac_2016_sedts},
$l\leq|\Omega(\pmc{l}{i})|\leq4l$. To guarantee the number of paths
selected by $AT$ is not larger than list size, $4l\leq\mathcal{L}$
and $AT=\pmc{\mathcal{L}/4}{i}$.}, meaning that only $\mathcal{L}/4$ PMs are guaranteed to be better
than $AT$ in this condition. The other $3\mathcal{L}/4$ paths will
be selected randomly. Intuitively, the lower bound of the performance
of this method is that of an LSCD with an equivalent list size of
$\mathcal{L}/4$. If we consider more unreliable bits in an MBD, the
error correction performance will be further degraded. In this sub-section,
we will propose a method to extend the original DTS to solve this
issue. 

We first consider the decoding of a $T$-bit tuple whose bits are
the leaf nodes of a sub-tree of the scheduling tree and the tuple
contains exactly one unreliable bit. We call this a \define{single-unreliable-bit tuple}
(SUBT). Later we will extend the idea to tuples that have different
combinations of bits. For a SUBT, other than the unreliable bit, each
bit is either a frozen bit or a reliable bit. As only the unreliable
bit needs to be expanded, there are only two expanded paths for each
SUBT and hence the PMU equations in \eqref{eq:mbd_pmu} are modified
as
\begin{equation}
\begin{cases}
\pmx{2l}{i+T} & =\pmc{l}{i}+\underbrace{{\textstyle \sum_{j=0}^{T-1}}(\alpha_{j}^{l}\oplus\Theta((\llr{t}{j})^{l}))\cdot|(\llr{t}{j})^{l}|}_{\Delta_{A}},\\
\pmx{2l+1}{i+T} & =\pmc{l}{i}+\underbrace{{\textstyle \sum_{j=0}^{T-1}}(\beta_{j}^{l}\oplus\Theta((\llr{t}{j})^{l}))\cdot|(\llr{t}{j})^{l}|}_{\Delta_{B}},
\end{cases}\label{eq:subt_pmu}
\end{equation}
where the $L_{j}^{t}$s are the LLR inputs at stage $t$ ($t=\log_{2}T$),
and $\alpha_{0}^{l}\alpha_{1}^{l}...\alpha_{T-1}^{l}=A^{l}$ and $\beta_{0}^{l}\beta_{1}^{l}...\beta_{T-1}^{l}=B^{l}$
are the encoded sub-vectors $V$ at stage $t$ when the only unreliable
bit in the SUBT is assumed to be 0 and 1, respectively. It can be
seen that non-zero penalties, denoted as $\Delta_{A}$ and $\Delta_{B}$
(the second terms at the right hand side of \eqref{eq:subt_pmu}),
are added for the updating of both PMs, and it is different from the
PMU of the traditional LSCD in \eqref{eq:pmu_aprx} where the penalty
is added to only one PM. If all the other bits are frozen bits (we
denote this SUBT as a rate-$\frac{1}{T}$ tuple), $A^{l}$ and $B^{l}$
are unique for path $l$ and can be obtained by \eqref{eq:mbd_evl}.
If there are some reliable bits, $A^{l}$ and $B^{l}$ have multiple
candidates and can be obtained by 
\begin{equation}
\begin{cases}
A^{l} & =\underset{V\in\mathbf{V0}}{\operatorname{arg\,min}}\{\sum_{j=0}^{T-1}(v_{j}\oplus\Theta((\llr{t}{j})^{l}))\cdot|(\llr{t}{j})^{l}|\},\\
B^{l} & =\underset{V\in\mathbf{V1}}{\operatorname{arg\,min}}\{\sum_{j=0}^{T-1}(v_{j}\oplus\Theta((\llr{t}{j})^{l}))\cdot|(\llr{t}{j})^{l}|\},
\end{cases}\label{eq:subt_evl}
\end{equation}
where $\mathbf{V0}$ and $\mathbf{V1}$ are two sets which include
all the possible combinations of $V$ when the unreliable bit in the
SUBT is assumed to be 0 and 1, respectively. \define{Maximum-likelihood detections}
(MLDs) on $\mathbf{V0}$ and $\mathbf{V1}$ is required to ensure
the selected $A^{l}$ and $B^{l}$ have the smallest $\Delta_{A}$
and $\Delta_{B}$ on the original PMs. Correspondingly, the $T$-bit
decoded vectors of the two expanded paths can be obtained by $A^{l}\cdot\genmat{t}$
and $B^{l}\cdot\genmat{t}$, respectively.

As the path expansion according to \eqref{eq:subt_pmu} and \eqref{eq:subt_evl}
generate $2\mathcal{L}$ expanded paths, DTS can be used for a SUBT.
Comparing this method with single bit decoding using the original
DTS and selective expansion, the error correction performance will
not be degraded. This is because the MLD in \eqref{eq:subt_evl} guarantees
any survival path kept by the DTS for a SUBT will not have a larger
PM than the one obtained by the original DTS if they are from the
same path before the expansion and have an identical value for the
unreliable bit.

\subsection{Latency Analysis of DTS for a SUBT \label{subsec:lt_an_opt}}

\begin{figure}
\includegraphics[width=8.8cm]{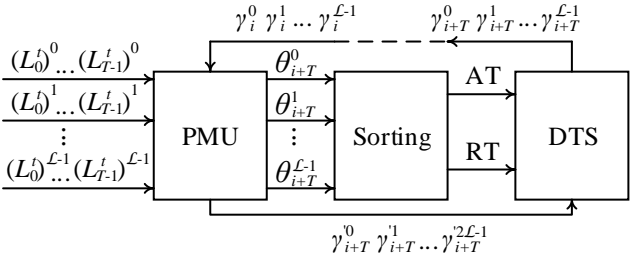}

\caption{Block diagram of the datapath of a DTS for a SUBT.}

\label{fig:dts4subt}
\end{figure}

For hardware implementation, we need multiple cycles to execute DTS
for a SUBT. The block diagram of the datapath is shown in \figref
\ref{fig:dts4subt}. Each block represents a step in the DTS for a
SUBT and is assumed to take one cycle. First, the PMU is calculated
based on \eqref{eq:subt_pmu} and \eqref{eq:subt_evl}. The inputs
are PMs of the existing survival paths, $\pmc{l}{i}$s, and LLRs from
stage $t$ of the $\mathcal{L}$ SCDs, $(\llr{t}{j})^{l}$s, where
$l\in[0,\mathcal{L}-1]$ and $j\in[0,T-1]$. The outputs are $2\mathcal{L}$
expanded PMs, $\pmx{k}{i+T}$s, where $k\in[0,2\mathcal{L}-1]$. In
the second cycle, the thresholds, $AT$ and $RT$, are extracted according
to \eqref{eq:atrt} based on the sorting result of the $\mathcal{L}$
$\theta_{i+T}^{l}$s, where $\theta_{i+T}^{l}=\text{min}\{\gamma_{i+T}^{'2l},\gamma_{i+T}^{'2l+1}\}$.
The proof of the validity of the thresholds is the same as that of
the traditional DTS in \cite{yzfan_hkust_jsac_2016_sedts} and will
not be discussed here. It is noted that in the original DTS, $\theta_{i+T}^{l}=\gamma_{i+1}^{'2l}=\gamma_{i}^{l}$
as penalty only added for the term $\gamma_{i+1}^{'2l+1}$ and hence
$\gamma_{i+1}^{'2l}$ is always smaller than $\gamma_{i+1}^{'2l+1}$.
Thus the threshold can be calculated based on the $\mathcal{L}$ $\gamma_{i}^{l}$s
which are known right after the decoding of bit $i$ and the sorting
of these $\mathcal{L}$ $\gamma_{i}^{l}$s can be done in parallel
with the LLR computation in the SCD to hide the sorting latency. However,
in DTS for a SUBT, the penalty values can be added in the computation
of both $\gamma_{i+1}^{'2l}$ and $\gamma_{i+1}^{'2l+1}$, so $\theta_{i+T}^{l}$
and hence the threshold values cannot be computed until the PMU generates
the result in the first cycle and sorting is needed in the second
cycle to obtain the threshold values. Finally, a DTS operation is
executed to select the paths to be kept based on the extracted $AT$
and $RT$ in the last cycle. In summary, the latency of one DTS for
a SUBT is three cycles. Of the three operations, the time delay due
to the MLDs in the PMU block is the highest and dictates the clock
frequency. We will discuss how to solve this problem in \secref \ref{sec:arch_llls}.

\subsection{Special Multi-bit Patterns \label{subsec:special_mbptrn}}

In some special multi-bit patterns, the three-cycle latency can be
reduced because some of the steps are not required as described below.
For simplicity, we denote these special patterns as SP1 and SP2, respectively. 

\subsubsection*{A SUBT with no frozen bit (SP1)}

If a SUBT does not include any frozen bit, it means all the bits apart
from the unreliable bit are reliable and all $2^{T}$ combinations
of $V$ are possible candidates of the encoded tuple in \eqref{eq:subt_evl},
causing a large computational load for the MLDs. According to the
properties of polar codes and selective expansion \cite{yzfan_hkust_jsac_2016_sedts,bli_huawei_istc_2014_hybrid},
the unreliable bit in this kind of tuple has the lowest reliability
and it is always the first bit of such a tuple. Based on this property,
we propose the following scheme to decode such a tuple. Specifically,
we use the original DTS \cite{yzfan_hkust_jsac_2016_sedts} to decode
the first bit. The magnitude of the LLR output at the first bit is
computed through a series of F-functions and hence its value $|(L_{0}^{0})^{l}|$
is equal to the minimum of all the LLR magnitudes at the root of the
tuple at stage $t$ which is denoted as $|(L_{k}^{t})^{l}|$, where
$k=\underset{j}{\operatorname{arg\,min}}|(L_{j}^{t})^{l}|$. Hence
the penalty values for the two paths expanded from this unreliable
bits are 0 and $|(L_{k}^{t})^{l}|$, respectively. The rest of the
reliable bits do not contribute any penalties to the PMs as no path
will be expanded from them.

\begin{figure}
\subfloat[]{\includegraphics[width=8.8cm]{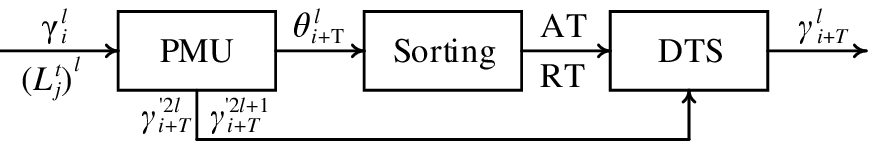}

\label{subfig:rd_lt_subt}}

\subfloat[]{\includegraphics[width=8.8cm]{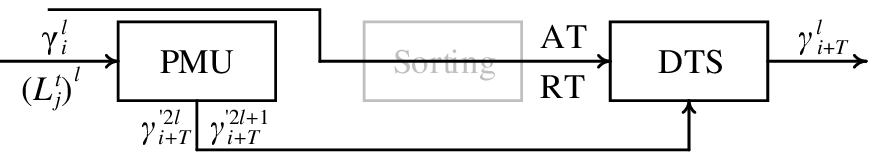}

\label{subfig:rd_lt_subt_nofrz}}

\subfloat[]{\includegraphics[width=8.8cm]{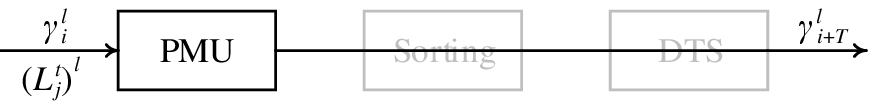}

\label{subfig:rd_lt_nubt}}

\caption{Original schedule for (a) a SUBT, modified schedule for (b) a tuple
with no frozen bit, and (c) an tuple with no unreliable bit.}
\label{fig:rd_lt}
\end{figure}

Next we need to find the decoded sub-vector of the reliable bits to
fill in the rest of the bits of the two expanded paths for this SUBT.
To decode these reliable bits, we can treat each of the two expanded
paths as a single-parity-check code \cite{gsarkis_mcgill_jsac_2014_fast}
in the traditional SCD with the unreliable bit as its parity bit (one
path with the parity bit equal to 0 and the other with the parity
bit equal to 1). As discussed in \cite{gsarkis_mcgill_jsac_2014_fast},
the parity bits of $A^{l}$ and $B^{l}$ are calculated as
\begin{equation}
\begin{array}{cc}
\eta_{A}={\displaystyle \sum_{j=0}^{T-1}}\Theta((L_{j}^{t})^{l}), & \eta_{B}=1+{\displaystyle \sum_{j=0}^{T-1}}\Theta((L_{j}^{t})^{l}),\end{array}\label{eq:parity}
\end{equation}
respectively, and then $A^{l}$ and $B^{l}$ can be obtained as
\begin{equation}
\begin{cases}
A^{l}=[\Theta((L_{0}^{t})^{l}),..,\Theta((L_{k}^{t})^{l})+\eta_{A},..,\Theta((L_{T-1}^{t})^{l})],\\
B^{l}=[\Theta((L_{0}^{t})^{l}),..,\Theta((L_{k}^{t})^{l})+\eta_{B},..,\Theta((L_{T-1}^{t})^{l})].
\end{cases}\label{eq:hd_parity}
\end{equation}
The decoded sub-vectors can then be obtained by encoding $A^{l}$
and $B^{l}$. This method expands two paths with the smallest penalty
values, 0 and $|(L_{k}^{t})^{l}|$, so the error correction performance
is guaranteed to be not worse than the results get by \eqref{eq:subt_pmu}
and \eqref{eq:subt_evl}. The computational complexity is significantly
reduced as we do not need to execute MLD to obtain $A^{l}$ and $B^{l}$.
Moreover, as one of the two expanded paths does not have a penalty
and keeps its original PM value, the threshold values can be pre-extracted
similar to the original DTS and the extra sorting cycle can be hidden
into the SCD computation cycle. Thus one cycle is saved as shown in
\figref \ref{fig:rd_lt}(b). 

\subsubsection*{A tuple with only frozen bits or only reliable bits (SP2)}

In these two multi-bit patterns, there will be no path expansions
and hence no DTS is needed. According to \tabref \ref{tab:se}, for
a tuple with only frozen bits, a PMU needs to be executed and the
decoded bits are all zero; for a tuple with only reliable bits, the
penalty values of all the paths are zero and the decoded bits are
$[\Theta((L_{0}^{t})^{l}),..,\Theta((L_{T-1}^{t})^{l})]\cdot\genmat{t}$.
Consequently, two cycles are saved for both patterns, as shown in
\figref \ref{fig:rd_lt}(c). 

These special patterns have at most only one unreliable bit so they
are treated as SUBTs in the following. It is also noted that these
special patterns are similar to the four kinds of special codes in
the fast-SSC-based algorithms. Specifically, an SP2 with only frozen
bits is a rate-0 code. An SP1 and an SP2 with only reliable bits are
rate-1 codes. However, the decoding algorithm of rate-1 code still
requires a few path expansions, while our method only requires at
most one path expansion and thus fewer clock cycles are needed for
LM.

\subsection{Multi-bit Double Thresholding Scheme\label{subsec:mb_dts}}

\begin{figure}
\subfloat[]{\includegraphics[width=4.4cm]{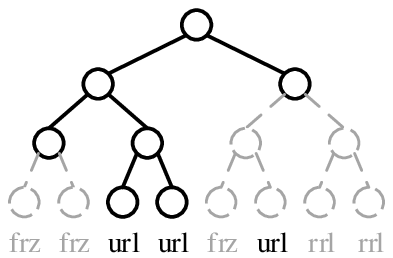}

\label{subfig:tree_tw_dts}}\subfloat[]{\includegraphics[width=4.4cm]{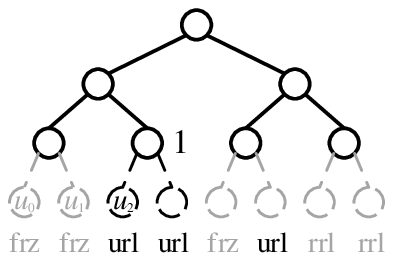}

\label{subfig:tree_mb_dts}}

\caption{Trimmed scheduling trees when MB-DTS is applied for (a) the whole
tree, and (b) sub-trees rooted at stage 1. 'frz', 'url' and 'rrl'
represent frozen, unreliable and reliable bits, respectively.}

\label{fig:trimming}
\end{figure}

We have discussed how to apply the DTS for decoding multi-bit patterns.
For a block of polar code that contains many unreliable bits, we first
present a tuple dividing scheme to separate a block of code into multiple
SUBTs. Based on this, an MB-DTS is proposed and its decoding latency
and complexity are discussed.
\begin{algorithm}
\caption{Tuple dividing scheme for a block of polar code}

\label{alg:1}

[Divided\_tuples,$N_T$]=TUPLE\_DIV(Root)\\
\Begin{	
	\If{\textrm{Root} is a SUBT}{	
		Trimming the sub-tree\;
		\Return{[Root,1]}\;
	}
	\Else{
		[Tuples$_{\text{left}}$,$N_{\text{left}}$]=TUPLE\_DIV(Child$_{\text{left}}$)\;
		[Tuples$_{\text{right}}$,$N_{\text{right}}$]=TUPLE\_DIV(Child$_{\text{right}}$)\;
		\Return{[(Tuples$_{\text{left}}$, Tuples$_{\text{right}}$),$N_{\text{left}}+N_{\text{right}}$]}\;
	}
}
\end{algorithm}

The tuple dividing scheme is summarized in \algref \ref{alg:1}.
The main function ``TUPLE\_DIV'' takes a tree as input and returns
a set of all the divided SUBTs and its cardinality, $N_{T}$. It is
recursively applied on the scheduling tree to divide it into SUBTs. 

The proposed MB-DTS first divides the whole scheduling tree according
to the tuple dividing scheme and then applies DTS for the leaf nodes
of the trimmed tree, which correspond to the SUBTs. An example is
shown in \figref \ref{fig:trimming}(a). Each leaf node in the trimmed
scheduling tree represents a SUBT which can be decoded by the DTS
in three cycles. If the tuple is an SP1 or SP2, one or two cycles
can be saved, respectively. The LLR calculations in the trimmed scheduling
tree are executed according to the depth-first traversal schedule,
and the latency of each node is one cycle. Consequently, the total
latency of traversing such a trimmed scheduling tree is given by
\begin{equation}
\mathcal{D_{\text{MB-DTS}}}=\mathcal{N}_{\text{node}}+3\cdot\mathcal{N}_{\text{leaf}}-\mathcal{N}_{\text{SP1}}-2\cdot\mathcal{N}_{\text{SP2}},\label{eq:lat_mb_dts}
\end{equation}
where $\mathcal{N}_{\text{node}}$ is the total number of the nodes
(except the root node), $\mathcal{N}_{\text{leaf}}$ is the number
of leaf nodes in this tree, and $\mathcal{N}_{\text{SP1}}$ and $\mathcal{N}_{\text{SP2}}$
are the numbers of leaf nodes corresponding to SP1 and SP2, respectively.
In the example shown in \figref \ref{fig:trimming}(a), $\mathcal{N}_{\text{node}}=6$,
$\mathcal{N}_{\text{leaf}}=4$, $\mathcal{N}_{\text{SP1}}=2$ (a single
unreliable bit is regarded as an SP1 here) and $\mathcal{N}_{\text{SP2}}=1$,
so the total latency is $\mathcal{D_{\text{MB-DTS}}}=14$ cycles.
It can be seen that the latency is greatly influenced by $\mathcal{N}_{\text{leaf}}$,
which depends on the number of unreliable bits. Thus the selective
expansion algorithm helps to greatly reduce the latency by reducing
the number of bits which need path expansions.

One of the issues of the MB-DTS for LSCD with a large list size is
that the size of the multi-bit tuples is not fixed. It is not favorable
for a regular hardware implementation. Also, if the tuple size is
too large, the PMU operations will be too complex and the corresponding
critical path delay will be high. To solve these problems, we divide
the scheduling tree into two parts at stage $m$ and only apply MB-DTS
to sub-trees rooted at stage $m$, i.e., DTS is only applied for leaf
nodes at the stages lower than $m$ in a trimmed tree. This restricts
the size of the tuple to a maximum length of $T_{\text{max}}=2^{m}=M$
so that the computational complexity and critical path delay can be
bound to a reasonable value. For stages higher than or equal to $m$,
LLR calculations are executed according to the traditional LSCD schedule.
\figref \ref{fig:trimming}(b) shows an example of an MB-DTS with
$M=2$. MB-DTS is only applied to the 2-bit sub-trees rooted at stage
1. Tuple dividing scheme is only applied to node 1 which includes
two unreliable bits. 

\section{Partial G-node Look-ahead\label{sec:pglah}}

In this section, we will present a partial G-node look-ahead scheme
to reduce the latency of the SCD computation for the LSCD with a large
list size. First, the G-node look-ahead scheme proposed in \cite{czhang_umn_icc_2012_lookahead}
for the conventional SCD will be reviewed. Then we will show how to
modify and apply this scheme to a semi-parallel LSCD architecture
to reduce the latency while keeping the hardware overhead to a minimum.

\subsection{Review of G-node Look-ahead}

In traditional SCD computation, G-node calculations are executed according
to \eqref{eq:g_func} after the partial-sums are generated, which
takes one clock cycle. If the dependency on the partial-sum is removed,
the G-nodes can be calculated at the same time with the F-nodes at
the same stage and the extra cycle is saved. In the GLAH scheme \cite{czhang_umn_icc_2012_lookahead},
the dependency is removed by unconditionally computing the G-node
twice assuming the partial-sums to be 0 and 1. Both results are stored
temporarily and the correct one will be selected directly when the
actual partial-sum is generated later. As half of the nodes in the
scheduling tree are G-nodes, the overall latency of SCD can be reduced
by half. The saving in latency comes with the cost of extra computations
and memory storage. For every two input LLRs, three output LLRs need
to be calculated and stored at the same time, one for the F-node and
two for the pre-computed G-node. The memory usage is hence about three
times that of the traditional SCD \cite{czhang_umn_icc_2012_lookahead}
and the cost is particularly high for LSCD with a large list size
as there are $\mathcal{L}$ SCDs.

In the rest of this section, we will discuss how we can use GLAH on
LSCD for the most latency-saving while keeping the hardware and memory
overhead to a minimum. 

\subsection{Partial G-node Look-ahead\label{subsec:pglah_parallelism}}

\begin{figure}
\includegraphics[width=8.8cm]{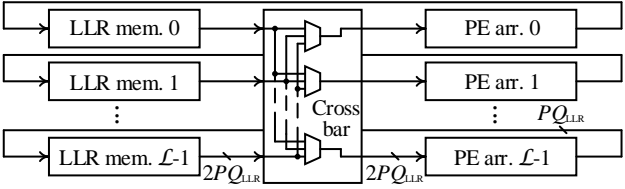}\caption{Block diagram of the SCD computation in an LSCD architecture.}

\label{fig:llr_loop}
\end{figure}
\begin{table*}
\caption{Relationship of parallelism, latency and memory requirement with stage
index in a semi-parallel LSCD architecture}

\centering{}

{\small{}\setlength{\tabcolsep}{5pt}\setlength{\extrarowheight}{1pt}}{\small \par}

\begin{tabular}{>{\centering}m{1.5cm}|>{\centering}m{1.5cm}|>{\centering}m{1.5cm}|>{\centering}m{1.5cm}|>{\centering}m{1.5cm}|>{\centering}m{1.5cm}|>{\centering}m{1.5cm}|>{\centering}m{1.5cm}|>{\centering}m{1.5cm}}
\hline 
 & {\small{}Stage} & {\small{}Parallel-} & {\small{}Latency} & \multicolumn{2}{c|}{{\small{}Traditional}} & \multicolumn{2}{c|}{{\small{}P-GLAH}} & {\small{}\# of words}\tabularnewline
\cline{5-8} 
 & {\small{}index} & {\small{}ism} & {\small{}(cycles)} & {\small{}Mem. bits} & {\small{}Data BW.} & {\small{}Mem. bits} & {\small{}Data BW.} & {\small{}($2PQ_{\text{LLR}}$)}\tabularnewline
\hline 
\hline 
 & {\small{}channel} & {\small{}-} & {\small{}-} & {\small{}$NQ_{\text{LLR}}$} & {\small{}-} & {\small{}$NQ_{\text{LLR}}$} & {\small{}-} & {\small{}-}\tabularnewline
\cline{2-9} 
{\small{}Semi-} & {\small{}$n-1$} & {\small{}$N/2$} & {\small{}$N/2P$} & {\small{}$NQ_{\text{LLR}}/2$} & {\small{}$2PQ_{\text{LLR}}$} & {\small{}$NQ_{\text{LLR}}/2$} & {\small{}$2PQ_{\text{LLR}}$} & {\small{}$N/4P$}\tabularnewline
\cline{2-9} 
{\small{}parallel} & {\small{}...} & {\small{}...} & {\small{}...} & {\small{}...} & {\small{}...} & {\small{}...} & {\small{}...} & {\small{}...}\tabularnewline
\cline{2-9} 
 & {\small{}$p+1$} & {\small{}$2P$} & {\small{}2} & {\small{}$2PQ_{\text{LLR}}$} & {\small{}$2PQ_{\text{LLR}}$} & {\small{}$2PQ_{\text{LLR}}$} & {\small{}$2PQ_{\text{LLR}}$} & {\small{}1}\tabularnewline
\hline 
\hline 
 & {\small{}$p$} & {\small{}$P$} & {\small{}1} & {\small{}$PQ_{\text{LLR}}$} & {\small{}$2PQ_{\text{LLR}}$} & {\small{}$2PQ_{\text{LLR}}$} & {\small{}$2PQ_{\text{LLR}}$} & {\small{}1}\tabularnewline
\cline{2-9} 
{\small{}Fully-} & {\small{}$p-1$} & {\small{}$P/2$} & {\small{}1} & {\small{}$PQ_{\text{LLR}}/2$} & {\small{}$PQ_{\text{LLR}}$} & {\small{}$PQ_{\text{LLR}}$} & {\small{}$2PQ_{\text{LLR}}$} & {\small{}1}\tabularnewline
\cline{2-9} 
{\small{}parallel} & {\small{}...} & {\small{}...} & {\small{}...} & {\small{}...} & {\small{}...} & {\small{}...} & {\small{}...} & {\small{}...}\tabularnewline
\cline{2-9} 
 & {\small{}0} & {\small{}1} & {\small{}1} & {\small{}$Q_{\text{LLR}}$} & {\small{}$2Q_{\text{LLR}}$} & {\small{}$2Q_{\text{LLR}}$} & {\small{}$4Q_{\text{LLR}}$} & {\small{}1}\tabularnewline
\hline 
\end{tabular}

\label{tab:parallelism}
\end{table*}
For an efficient hardware implementation, a semi-parallel architecture
\cite{cleroux_mcgill_tsp_2013_semiparallel} is usually used in a
traditional LSCD architecture \cite{abalatsoukas_epfl_tsp_2015_llrlscd,yzfan_hkust_jsac_2016_sedts}.
The block diagram of the SCD computation in the LSCD architecture
is shown in \figref \ref{fig:llr_loop}, which contains $\mathcal{L}$
blocks of LLR memory, $\mathcal{L}$ \define{processing element}
(PE) arrays and one $\mathcal{L}\times\mathcal{L}$ crossbar. The
memories are used to store the intermediate LLRs at each stage. Each
PE array uses $P=2^{p}\,(\ll N)$ PEs for the SCD computation. After
an LM, the LLR values of some paths need to be copied to the other
LLR memory of other SCDs to continue the subsequent decoding operations
if both of the expanded paths are kept. To eliminate the costly data
movement between the memories, usually a pointer-based updating mechanism
is used instead \cite{ital_ucsd_tit_2015_list}. To support this pointer-based
operation, the crossbar is used to align the data stored in the memories
and the PE arrays for correct operations \cite{abalatsoukas_epfl_tcasii_2014_archlscd,abalatsoukas_epfl_tsp_2015_llrlscd}.
The bit-width of the read port of each memory and the crossbar is
$2PQ_{\text{LLR}}$ bits, where $Q_{\text{LLR}}$ is the number of
quantization bits for the LLR operands. If GLAH is used in this architecture,
similar as the architecture presented in \cite{czhang_umn_icc_2012_lookahead},
the size of the memories is tripled and the data moved from the memories
to PE arrays are doubled to $4PQ_{\text{LLR}}$ bits because each
LLR input of the G-function has two candidates. This increases the
complexity of the memories and the crossbar. 

\tabref \ref{tab:parallelism} summarizes the relationships of the
computational parallelism (in terms of the number of F- or G-functions
that are executed in parallel), the latency, the memory storage requirement,
the data bandwidth requirement and the number of memory words with
the stage index for a semi-parallel LSCD architecture that has $P$
PEs for each path. At stage $s$, the decoding parallelism is $2^{s}$.
According to these relationships, we can separate all the stages into
two groups:
\begin{enumerate}
\item Stages with indices smaller than or equal to $p$ are called fully-parallel
stages, as their parallelism is not larger than $P$ and the computations
can be finished in one cycle.
\item Stages with indices larger than $p$ are called semi-parallel stages,
as their parallelism is larger than $P$ and the computations take
multiple cycles.
\end{enumerate}
Based on the above analysis, we propose a partial G-node look-ahead
scheme. Specifically, for the fully-parallel stages, we use the GLAH
scheme; and for the semi-parallel stages, we just use the traditional
decoding scheme without the GLAH.

Though the data bandwidth at the fully-parallel stages (except stage
$p$, which gets $2P$ LLR operands from the semi-parallel stages)
is doubled because of the GLAH calculation, it does not exceed the
maximum bandwidth requirement ($2PQ_{\text{LLR}}$ bits) and hence
is unchanged. The overall decoding latency is nearly halved because
most of the nodes in the scheduling tree belong to the fully-parallel
stages. Specifically, the latency of P-GLAH is given by
\begin{align}
\mathcal{D}_{\text{P-GLAH}} & =\text{\ensuremath{\mathcal{D}}}_{\text{trd}}-\text{\ensuremath{\Delta\mathcal{D}}}\label{eq:pglah}\\
 & =(2N+\frac{N}{P}\log_{2}(\frac{N}{4P}))-(N-\frac{N}{2P})\label{eq:pglah-2}\\
 & =N+\frac{N}{2P}+\frac{N}{P}\log_{2}(\frac{N}{4P}).\label{eq:pglah-final}
\end{align}
where $\text{\ensuremath{\mathcal{D}}}_{\text{trd}}$ is the latency
of the traditional semi-parallel architecture without GLAH computation
\cite{cleroux_mcgill_tsp_2013_semiparallel} and $\text{\ensuremath{\Delta\mathcal{D}}}$
is the latency saving which equals to the number of G-nodes in the
fully-parallel stages. For example, if $N=1024$ and $P=64$, the
latency of P-GLAH is about 51.2\% of that of the traditional SCD schedule,
$\mathcal{D}_{\text{trd}}$. This is very close to the latency of
the original GLAH, which is exactly 50\% of $\mathcal{D}_{\text{trd}}$. 

Next, we analyze the memory usage of the semi-parallel LSCD architecture
with P-GLAH. For the fully-parallel stages, the memory usage is $4PQ_{\text{LLR}}$
bits for each path, which equals to the sum of the memory bits of
all the stages shown in \tabref \ref{tab:parallelism}. It is noted
that, compared with the $2PQ_{\text{LLR}}$-bit memory usage in these
stages of the traditional LSCD, the required memory bits are only
doubled instead of tripled. This is because the calculation results
of the F-nodes will be used in the subsequent cycle and can be stored
in an extra $PQ_{\text{LLR}}$-bit bypass register bank\cite{cleroux_mcgill_tsp_2013_semiparallel}.
This also guarantees the $P$ pairs of GLAH calculations executed
at stage $p$ have a write data bandwidth of $2PQ_{\text{LLR}}$ bits
instead of $3PQ_{\text{LLR}}$ bits. For the semi-parallel stages,
as we do not use the GLAH scheme, the memory usage is the same as
that of the traditional architecture, which is $(\frac{N}{2P}-1)\cdot2PQ_{\text{LLR}}$
bits for each path. To support $\mathcal{L}$ parallelly-executed
SCDs, $\mathcal{L}$ copies of memories are needed for the fully-parallel
stages, the semi-parallel stages and the bypass memories. Besides,
$NQ_{\text{LLR}}$ memory bits are used to store the channel LLRs
which can be shared by all the paths. So, the overall size of the
LLR memory for an LSCD with list size equal to $\mathcal{L}$ is $[(\mathcal{L}+1)N+3\mathcal{L}P]\cdot Q_{\text{LLR}}$
bits. The size of the LLR memory is similar to that of the traditional
LSCD architecture such as \cite{yzfan_hkust_jsac_2016_sedts}. Specifically,
for the example mentioned in the last paragraph, the memory overhead
is only about 18\% for a list size of 32 and much smaller than the
overhead of the original GLAH \cite{czhang_umn_icc_2012_lookahead}. 

\section{High-Throughput Architecture for LSCD with A Large List Size \label{sec:arch_llls}}

In this section, we first present the overall architecture of the
proposed LSCD and analyze its hardware complexity. Then, we discuss
the details of the main sub-modules.

\subsection{Overall architecture\label{subsec:overall_arch}}

\begin{figure}
\includegraphics[width=8.8cm]{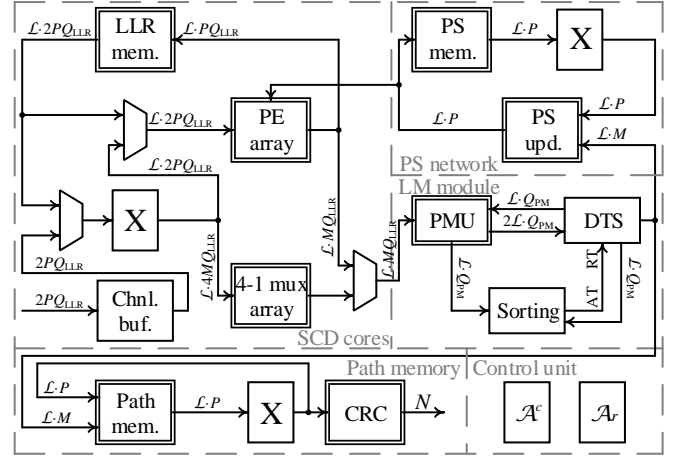}

\caption{The proposed high-throughput architecture for LSCD with a large list
size.}

\label{fig:blk_dgm_llslscd}
\end{figure}

The structure of the proposed architecture is shown in \figref \ref{fig:blk_dgm_llslscd}.
It consists of several modules: the SCD cores, the LM module, the
partial-sum network, the path memory and the control unit. To support
an LSCD with a list size equal to $\mathcal{L}$, most of the blocks
in \figref \ref{fig:blk_dgm_llslscd} are duplicated $\mathcal{L}$
times, and they are drawn with double-lined frames. %

The SCD cores are used to execute the SCD computations. The input
channel LLRs and the intermediate LLRs at each stage of SCD are stored
in the channel buffer and the LLR memories, respectively. Both of
the memories are implemented in words of $2PQ_{\text{LLR}}$ bits.
To save the memory usage, pre-computation memory-saving (PCMS) proposed
in \cite{crxiong_lehigh_tsp_2016_symbol} is applied, which means
the GLAH calculation is executed at the top stage $n-1$. These calculation
results are shared by all the $\mathcal{L}$ paths and hence channel
LLRs do not need to be stored once they are used in the first computation.
Thus only $\frac{3N}{2}$, instead of $\frac{\mathcal{L}N}{2}$, LLR
storage is needed at stage $n-1$. The crossbar selects the LLRs from
either the LLR memory or the channel buffer as input and sends the
data to the PE arrays for F- and G-node computation according to the
pointer-based lazy copy scheme \cite{abalatsoukas_epfl_tcasii_2014_archlscd}.
The PEs are designed to support both GLAH and non-GLAH computation,
and the detailed structure will be presented in \secref \ref{subsec:ppe}.
As MB-DTS is used, after the LLR calculations of stage $m$ is finished,
$M$ LLRs are sent to the LM module. Here, the schedule of the SCD
computations at stage $m$ and the LM operations are re-designed to
further reduce the latency, which will be detailed in \secref \ref{subsec:lat_ft}. 

The LM module directly implements the block diagram shown in \figref
\ref{fig:dts4subt}. The PMU block is used to calculate the PMs of
the expanded paths and the details will be presented in \secref \ref{subsec:pmu_str}.
The DTS block is used to realize the DTS algorithm, and a structure
similar to that presented in \cite{yzfan_hkust_jsac_2016_sedts} is
used. The sorter is used to sort the PMs of the $\mathcal{L}$ survival
paths in order to obtain the $AT$ and $RT$ values for the DTS for
the decoding of the next SUBT. These three blocks are mapped and connected
according to the schedule of decoding a SUBT as discussed in \secref
\ref{subsec:lt_an_opt} and \ref{subsec:special_mbptrn}.

The partial-sum network is used to update and store the partial-sums
required for the G-node computations. A folded partial-sum network
for semi-parallel SCD architecture similar to that proposed in \cite{yzfan_hkust_tsp_2014_effps}
is used for each path. The partial-sum memory and partial-sum update
block are duplicated $\mathcal{L}$ times to support $\mathcal{L}$
paths. Each copy of the partial-sum memory contains $N$ memory bits.
Among these bits, $P$ bits are stored in registers while the others
are stored in an SRAM whose port width equals to $P$ bits. After
MB-DTS for an $M$-bit tuple is finished, $\mathcal{L}\cdot P$ bits
of partial-sums from $\mathcal{L}$ paths are sent to the crossbar
for permutation according to the LM results. The permuted partial-sums
are updated with the $\mathcal{L}\cdot M$ decoded bits from the LM
module and then stored back to the partial-sum memory, while at the
same time sent to the PE arrays for G-node computations. More details
of the partial-sum network architecture can be found in \cite{yzfan_hkust_tsp_2014_effps}.

The path memory is used to store and update the partial decoded vectors
of each path. Its structure is similar to that of the partial-sum
network. $\mathcal{L}$ copies of path memories are used to store
the decoded bits of $\mathcal{L}$ paths and a crossbar is used to
update each path according to the LM results. The newly decoded bits
are appended to the partial decoded vectors and the updating block
is simply implemented using shifters. When the decoding of a code
block is finished, all the paths are checked with the CRC unit and
the one that passes the checking is selected as the output word.

The control unit generates the control signal for the decoder. The
frozen set $\mathcal{A}^{c}$ and the reliable set $\mathcal{A}_{r}$
are stored in a small ROM in this unit. Based on these sets, the size
of the tuples and the other related control signals are generated
on-line according to Algorithm 1. For the decoder flexibility, we
can simply change the contents in the ROM when the code sets are changed
and the hardware architecture does not need to be modified.

\subsection{Programmable Processing Element\label{subsec:ppe}}

\begin{figure}
\includegraphics[width=8.8cm]{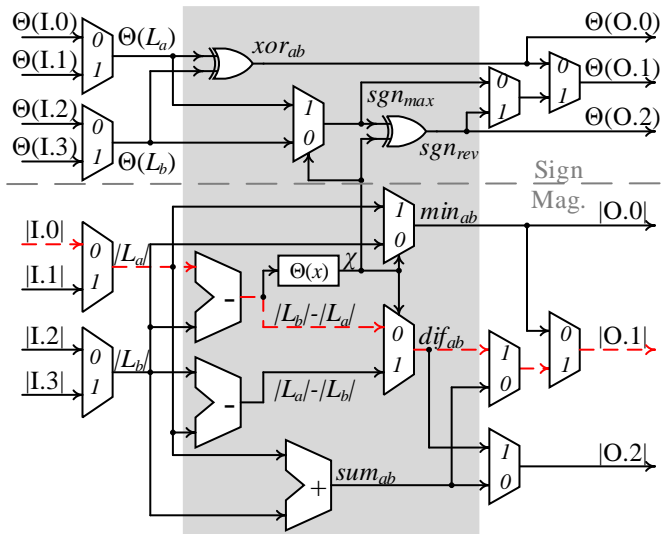}\caption{Structure of a programmable processing element.}

\label{fig:ppe}
\end{figure}
\begin{table}
\caption{Logic delay and complexity comparison (in terms of adders) of the
existing processing element structures}

\label{tab:pe}

\begin{tabular}{>{\centering}p{1.8cm}|c|c|c|c}
\hline 
\multicolumn{1}{>{\centering}p{1.8cm}}{} &  & {\small{}Proposed} & {\small{}\cite{cleroux_mcgill_tsp_2013_semiparallel}} & {\small{}\cite{czhang_umn_icc_2012_lookahead}}\tabularnewline
\hline 
\multicolumn{2}{c|}{{\small{}Is GLAH supported?}} & {\small{}$\checkmark$} &  & {\small{}$\checkmark$}\tabularnewline
\hline 
{\small{}Adders } & {\small{}on critical path} & {\small{}1} & {\small{}2} & {\small{}1(+2)}\tabularnewline
\cline{2-5} 
{\small{}(+converters)} & {\small{}in total} & {\small{}3} & {\small{}2} & {\small{}1(+5)}\tabularnewline
\hline 
\end{tabular}
\end{table}

As the P-GLAH scheme is used in the implementation of the LSCD, the
processing elements have to execute the GLAH and normal non-GLAH computations
at the fully-parallel and semi-parallel stages, respectively, and
hence a programmable processing element is required.

The structure of the programmable PE is shown in \figref \ref{fig:ppe}.
Sign-magnitude representation is used to represent the LLRs. The datapath
contains an input stage, a calculating stage and an output stage which
are described as below. 
\begin{itemize}
\item The input stage is configured according to the number of input candidates.
If there are only two input LLRs, i.e. when pre-computation is not
used, only two ports, I.0 and I.2, are activated and selected. Otherwise,
$L_{a}$ and $L_{b}$ are selected from the four pre-computed values
by the partial-sums. 
\item The calculating stage is used to generate the candidate values for
the output stage according to \eqref{eq:g_func} and \eqref{eq:min-sum}.
The datapath for the magnitudes mainly consists of three adders which
are used to calculate $|L_{a}|+|L_{b}|$, $|L_{a}|-|L_{b}|$ and $|L_{b}|-|L_{a}|$.
The overflow bit of $|L_{a}|-|L_{b}|$, marked as $\chi$, will be
used to select the minimum of the two magnitudes $min_{ab}$ as well
as the absolute value of the difference, $dif_{ab}$, from $\pm|L_{a}|\mp|L_{b}|$.
The $sum_{ab}$, $dif_{ab}$ and $min_{ab}$ are used as the candidates
at the output stage. Similarly, in the datapath for the sign bit,
three sign bits are generated. 
\item The output stage is configured according to whether GLAH is used or
not. There are three and one outputs when GLAH is and is not used,
respectively. O.1 will be sent to the memories under both conditions. 
\end{itemize}
The delay and complexity of the programmable PE is mainly brought
by the adders and the critical path includes only one stage of adder
and some multiplexers, which is highlighted with the dashed line in
\figref \ref{fig:ppe}. We compare the proposed programmable PE structure
with the existing ones from the literatures and the results are summarized
in \tabref \ref{tab:pe}. All the PE structures listed here use sign-magnitude
representation for the LLR operands. The PE structure in \cite{cleroux_mcgill_tsp_2013_semiparallel}
uses one fewer adder than ours; however, this is at the cost of larger
logic delay. Moreover, it cannot support the GLAH calculation. \cite{czhang_umn_icc_2012_lookahead}
uses an adder-subtractor to calculate multiple expressions for GLAH.
However, for such a structure, the representation of all the input
and output operands must be converted between 2's complement and sign-magnitude.
As a result, there are in total one adder and five converters (for
representation converting) and the critical path includes one stage
of adder and two stages of converters. It can be seen that the programmable
PE structure can realize the partial G-node look-ahead scheme with
low logic delay and moderate hardware complexity. 

\subsection{PMU Block in the LM Module\label{subsec:pmu_str}}

\begin{figure}
\includegraphics[width=8.8cm]{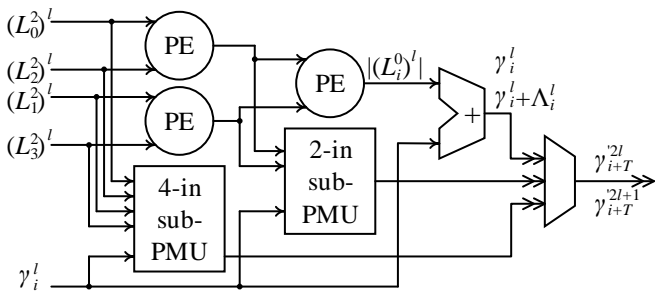}

\caption{Structure of the PMU block for an $M$-bit tuple in the MB-DTS ($M$=4).
An double arrow means the signal includes two updated PMs. }

\label{fig:pmu_structure}
\end{figure}
To implement the MB-DTS algorithm on hardware, the PMU block shown
in \figref \ref{fig:pmu_structure} is used in the LM module, which
directly implements the algorithm introduced in \secref \ref{subsec:mb_dts}. 

\begin{figure}
\includegraphics[width=8.8cm]{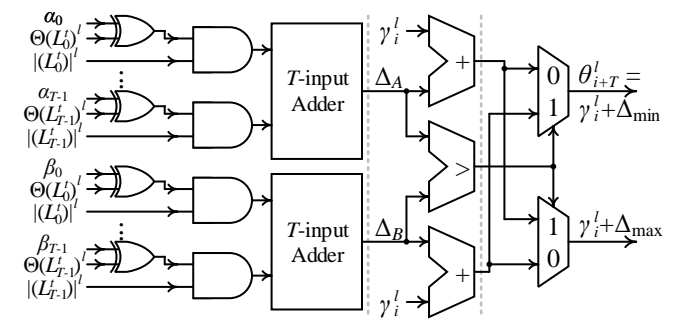}

\caption{Structure of a $T$-input sub-PMU block.}

\label{fig:sub_pmu}
\end{figure}

Let $m$ be the highest stage in the scheduling tree that MB-DTS can
be used. With the tuple dividing algorithm, the size of the tuple
decoded using MB-DTS $T$ ranges from 1 to $2^{m}$, indicating the
PMU block needs to update the PM with at most $M$ LLRs from one path.
In the example shown in \figref \ref{fig:pmu_structure}, $M=4$,
and it can be used to execute the PMU for tuples with $T=1,2\,\text{or}\:4$.
The inputs to the PMU are $M$ LLRs from stage $m$. If $T<M$, the
PMU cannot be executed immediately with the $M$ inputted LLRs. According
to \secref \ref{subsec:mb_dts}, $m-t$ stages of LLR calculations
are executed in $m-t$ cycles to obtain the valid LLRs for the PMU
of a $T$-bit tuple. Therefore, $m$ stages of programmable PEs are
needed to execute these LLR calculations and registers are used to
save the outputs at each stage. If $T=1$ or the tuple is an SP1,
the bit can be decoded with the original DTS. Consequently, the magnitude
value of the last stage of LLR calculation, $|(L_{i}^{0})^{l}|$,
which is the output of the last PE in \figref \ref{fig:pmu_structure},
is used as the penalty value $\Lambda_{i}^{l}$ in \eqref{eq:pmu_aprx}
and the PM values can be updated accordingly. Otherwise, if $T>1$,
after $m-t$ stages of LLR calculations, a $T$-input sub-PMU block
shown in \figref \ref{fig:sub_pmu} is used to update the PM values
for the $T$-bit tuple.

The $T$-input sub-PMU block is used to compute \eqref{eq:subt_pmu}
and \eqref{eq:subt_evl}. To implement the MLD in \eqref{eq:subt_evl},
we need to update up to $2^{T}$ PMs with penalty and find the two
minimum values out of two groups of $2^{T-1}$ values, which requires
$2^{T}$ $T$-input adders and two groups of $T-1$ stages of comparators,
respectively. To reduce the hardware complexity and time delay of
the datapath, in the real sub-PMU block, we do not execute MLD at
all. Instead, during the formation of the tuples, we further restrict
the tuple to be one of the following special patterns of SUBT: an
SP1, an SP2 or a rate-$\frac{1}{T}$ tuple. As discussed in \secref
\ref{sec:mbdts}, none of these tuple patterns require an MLD, so
the complex calculation of MLD is not required at all.

The structure of the $T$-input sub-PMU block is shown in \figref
\ref{fig:sub_pmu} and it consists of two identical sub-blocks, which
are used to calculate \eqref{eq:subt_pmu} with the corresponding
$A^{l}$ and $B^{l}$ in parallel. Both of the two sub-blocks are
activated when a rate-$\frac{1}{T}$ tuple is decoded, while only
one of them is activated when an SP2 is decoded. There are three stages
in the datapath, which are divided by the dotted lines in \figref
\ref{fig:sub_pmu}. The first stage consists of two $T$-input adders
to calculate the two penalty values, $\Delta_{A}$ and $\Delta_{B}$
in \eqref{eq:subt_pmu}, which are then sent to the second stage
to add up with the PM value of the current path. The last stage is
used to select the smaller of the the two updated PMs, $\theta_{i+T}^{l}$,
for the sorting process that may be executed in the next cycle.

The delay of the $T$-input sub-PMU block is mainly due to the adders
and comparators. A $t$-stage adder-tree is required to implement
the $T$-input adder. Thus, the critical path delay of the whole PMU
block lies at the $M$-input sub-PMU block which equals to that of
$m+1$ stages of adders. As mentioned in \secref \ref{subsec:mb_dts},
$M$ should not be too large so as to bound this critical path delay
to a moderate value. The datapaths to calculate other $T$-bit tuples
includes $m-t$ stages of programmable PEs (each has the delay of
about one adder) and $t+1$ stages of adders for the $T$-input sub-PMU
block and the total delay is also $m+1$ stages of adders. Thus, the
LLR calculations and PMU of any $T$-bit tuple can be merged and executed
in one cycle without increasing the critical path delay. If we consider
the latency to traverse a sub-tree at stage $m$ and below, the first
term in \eqref{eq:lat_mb_dts}, $\mathcal{N}_{\text{node}}$, can
be removed. We denote this modified MB-DTS with the optimized schedule
as \define{simplified MB-DTS} (SMB-DTS), and its latency in terms
of cycle can be expressed as%
\begin{equation}
\mathcal{D}_{\text{SMB-DTS}}=3\cdot\mathcal{N}_{\text{leaf}}-\mathcal{N}_{\text{SP1}}-2\cdot\mathcal{N}_{\text{SP2}}.\label{eq:lat_mb_dts-1}
\end{equation}

\subsection{Latency Fine-tuning by Datapath Optimization\label{subsec:lat_ft}}

\begin{figure}
\subfloat[]{\includegraphics[width=4cm]{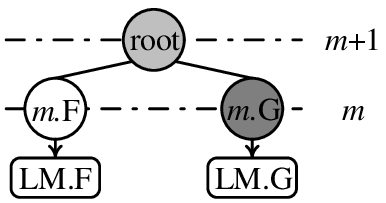}

}\subfloat[]{\includegraphics[width=4.7cm]{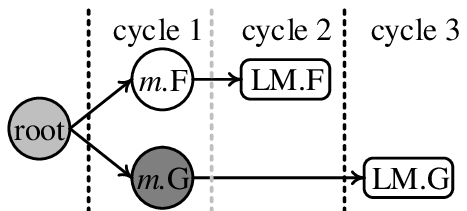}}\caption{(a) A sub-tree rooted at stage $m+1$, and (b) its schedule.}
\label{fig:subtree_mp1}
\end{figure}

\begin{figure}
\subfloat[]{\includegraphics[width=8.8cm]{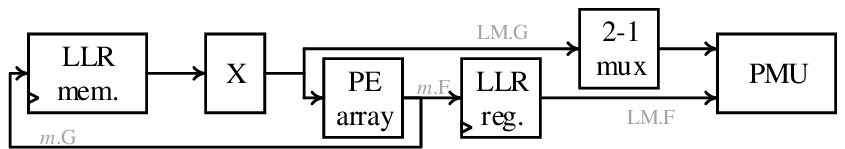}

}

\subfloat[]{\includegraphics[width=8.8cm]{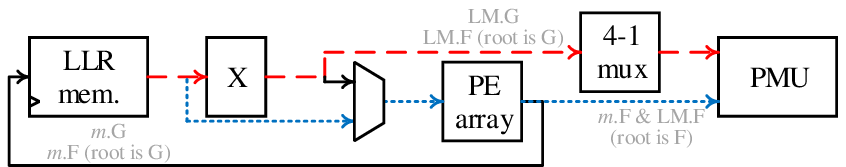}}\caption{Datapaths of the calculations near stage $m$ (a) by direct-mapping,
and (b) with the proposed optimization.}
\label{fig:arch_intf}
\end{figure}

In this sub-section, a critical path optimization scheme is proposed
to further reduce the latency. Let us take a snapshot of the decoding
process at the nodes near stage $m$ first. A sub-tree rooted at stage
$m+1$\footnote{Here, we assume that $p>m+1$ to ensure that GLAH is used for these
two stages. In real design, to bound the complexity of the PMU block,
$m$ is usually $\ll p$.} is shown in \figref \ref{fig:subtree_mp1}(a) and its decoding schedule
is shown in \figref \ref{fig:subtree_mp1}(b), in which the dotted
line represents a pipeline stage. In cycle 1, two LLR calculations
at stage $m$, $m$.F and $m$.G, are executed in parallel using GLAH.
The PMU of LM.F (the LM after $m$.F is finished) is executed in the
second cycle\footnote{Here, we assume an LM operation takes one cycle. If it takes multiple
cycles, the optimization scheme is still valid.}. After LM.F is finished, the pre-computed $m$.G results are selected
as the inputs of LM.G (the LM after $m$.G is finished) and the PMU
of LM.G is calculated immediately in cycle 3. An architecture that
directly mapped the above operations into hardware is shown in \figref
\ref{fig:arch_intf}(a). To reduce the decoding latency, we can remove
the pipeline stage between cycle 1 and 2 (the grey dotted line in
\figref \ref{fig:subtree_mp1}(b)). However, direct de-pipelining
makes the delay of this datapath (sum of the delay of the crossbar,
the PE array and the PMU) much longer than that of the other blocks
of the decoder, such as the DTS block, and affects the overall clock
frequency. By carefully analyzing the data dependency, we can optimize
the critical path of the de-pipelined datapath to fine-tune the latency
under the following two situations.
\begin{itemize}
\item If the root node at stage $m+1$ is a G-node, the inputs of $m$.F
are already pre-computed using GLAH several cycles earlier. This means
that we can also pre-compute the F-functions in $m$.F some cycles
before and we do not need to compute the LLR values in this cycle.
As each input LLR of $m$.F has two candidate values, there are four
possible combinations of the inputs and hence four possible outputs
of the $m$.F. We can calculate all these four F-functions when LM
operations of the previous leaf nodes are being executed. To do so,
we can re-use the programmable PEs in the SCD cores without adding
extra hardware as they are idle during the LM operation. A group of
4-1 mux arrays is needed to select the correct LLRs, as shown in \figref
\ref{fig:arch_intf}(b), and the total delay is the sum of the delay
of the crossbar, the mux and the PMU (highlighted with the dashed
line). 
\item If the root node at stage $m+1$ is an F-node, the inputs of $m$.F
are calculated at the last cycle. When an F-node is executed, the
data in the corresponding LLR memory of a PE array is always used
as its own input LLRs for $m$.F. Thus the crossbar is not needed
for this situation and can be bypassed. Hence, the architecture in
\figref \ref{fig:arch_intf}(a) can be de-pipelined as shown in \figref
\ref{fig:arch_intf}(b), and the total delay is the sum of the delay
of the mux, the PE array and the PMU (highlighted with the dotted
line). 
\end{itemize}
With the proposed latency fine-tuning scheme and the optimized datapath
shown in \figref \ref{fig:arch_intf}(b), the F-function at stage
$m$ and the PMU of its following LM can be executed in the same cycle
with a small critical path delay overhead. As there are $\frac{N}{2M}$
F-nodes on stage $m$, $\frac{N}{2M}$ cycles can be saved in total.%

\subsection{Decoding Latency of the Proposed LSCD Architecture}

Based on the above discussions, the overall latency of the LSCD architecture
with all the proposed schemes for decoding one frame is given by 
\begin{align}
\mathcal{D}_{\text{tot}} & =\underset{\text{LM}}{\underbrace{\Sigma(\mathcal{D}_{\text{SMB-DTS}})}}+\underset{\text{SCD}}{\underbrace{\mathcal{D}_{\text{P-GLAH}}}}-\underset{\text{Saving}}{\underbrace{(\mathcal{D}_{\text{fine}}+\mathcal{D}_{\text{zero}})}}.\label{eq:lat_tot}
\end{align}
Specifically, the first term represents the cycles required for decoding
the nodes at the stages lower than $m$, which equals the sum of the
cycles required for all the $M$-bit tuples decoded by SMB-DTS. $\mathcal{D}_{\text{SMB-DTS}}$
and $\mathcal{D_{\text{P-GLAH}}}$ can be calculated according to
\eqref{eq:lat_mb_dts-1} and \eqref{eq:pglah-final}, respectively.
$\mathcal{D}_{\text{fine}}=\frac{N}{2M}$ is the saving due to the
technique discussed in \secref \ref{subsec:lat_ft}. The last latency-saving
term $\mathcal{D}_{\text{zero}}$ is due to the structure of the polar
code and is described as follows.

For a block of polar code, the leaf node corresponding to the first
information bit in the scheduling tree is already known at the beginning
of the decoding. Specifically, all the bits before the first information
bit are frozen bits and the only partially-decoded vector is all-zero.
Thus, we can find the path from the root to the leaf node corresponding
to this information bit and calculate the nodes on this path. All
the partial-sums of the G-nodes on this path are zero, so there is
no data dependency. For example, in \figref \ref{fig:trimming}(b),
node 1 and hence the first information bit $u_{2}$ can be decoded
at the beginning of the decoding as $u_{0}$ and $u_{1}$ are frozen
and the partial-sums of node 1 are zeros. The decoding process for
the all-zero vector is not needed and the latency saved is denoted
as $\mathcal{D}_{\text{zero}}$.

For a conventional semi-parallel LSCD \cite{abalatsoukas_epfl_tsp_2015_llrlscd},
the baseline decoding latency is $2N$ cycles for the $\mathcal{L}$
parallel SCDs and $K$ cycles for LM. As the latency of our SMB-DTS
highly depends on the setting of the code, we will show the latency
saving by a numerical example in \secref \ref{subsec:lat_cal}.

\section{Experimental Results\label{sec:experiment}}

\subsection{Error Correction Performance of the Proposed Schemes\label{subsec:error_perf}}

To demonstrate the error correction performance of the proposed
schemes, simulations are done on the polar code of $(N,K,r)=(1024,512,24)$
over an AWGN channel.

\begin{figure*}
\subfloat[$M=2$]{\includegraphics[width=6cm]{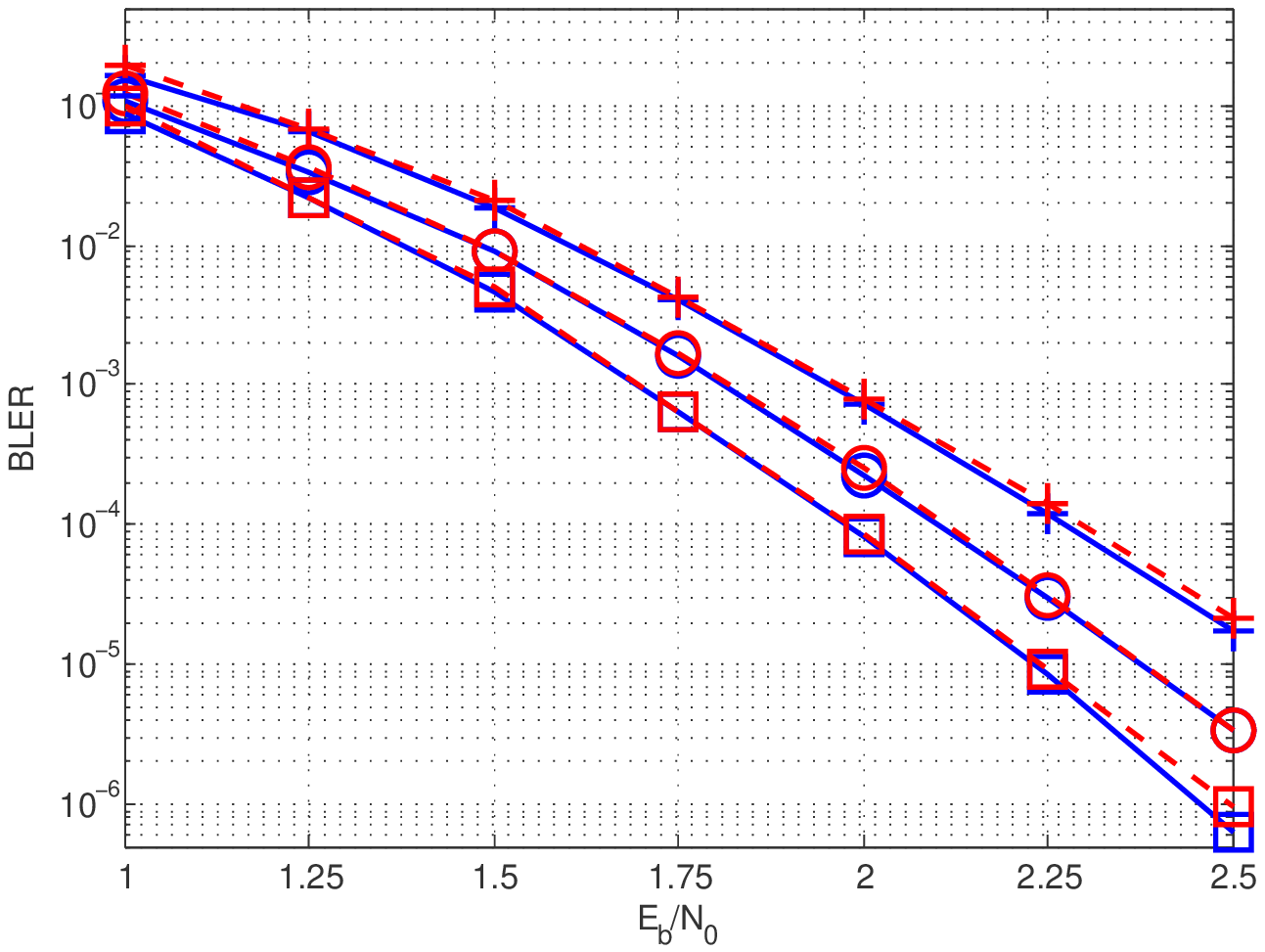}}\subfloat[$M=4$]{\includegraphics[width=6cm]{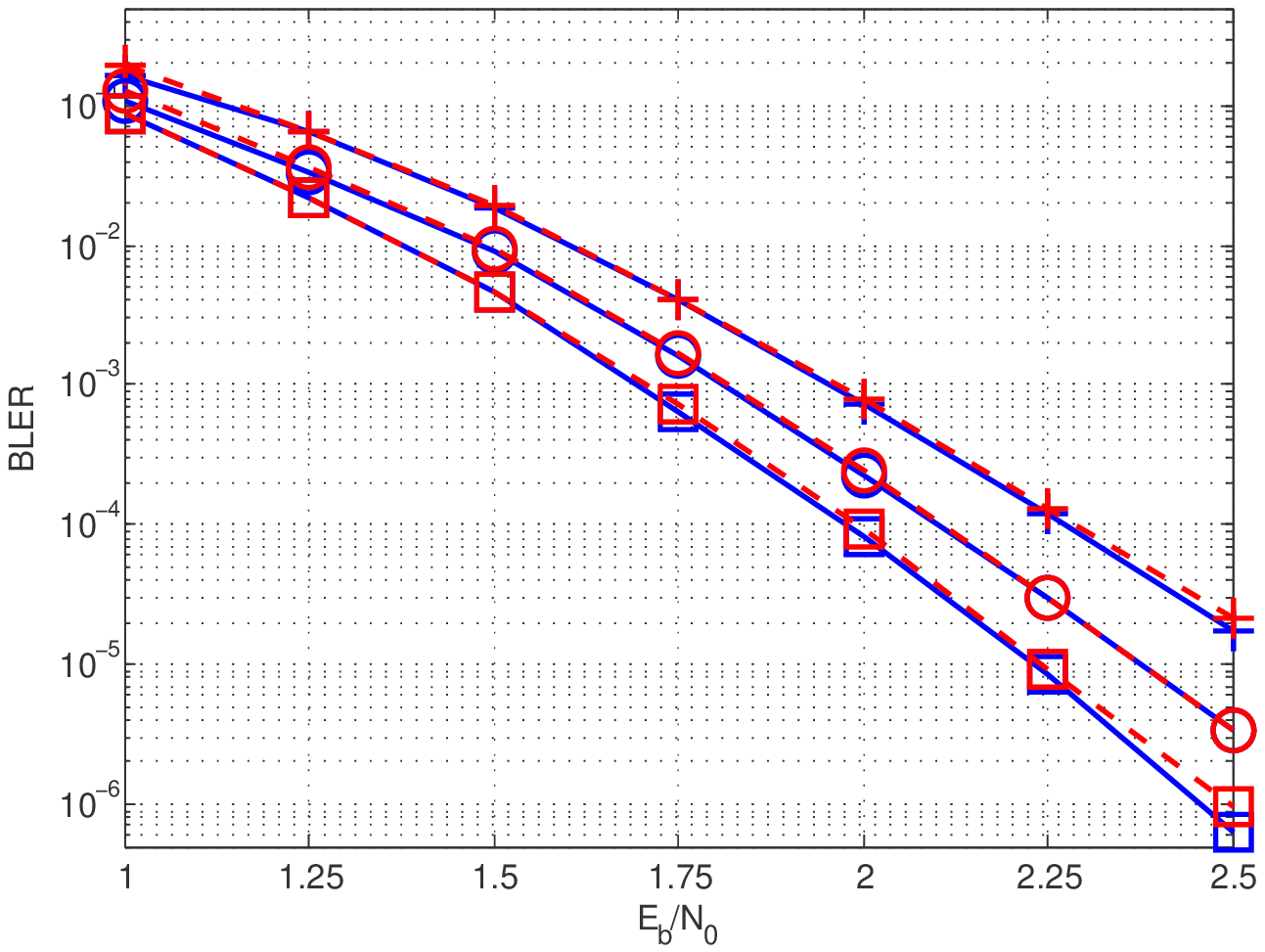}}\subfloat[$M=8$]{\includegraphics[width=6cm]{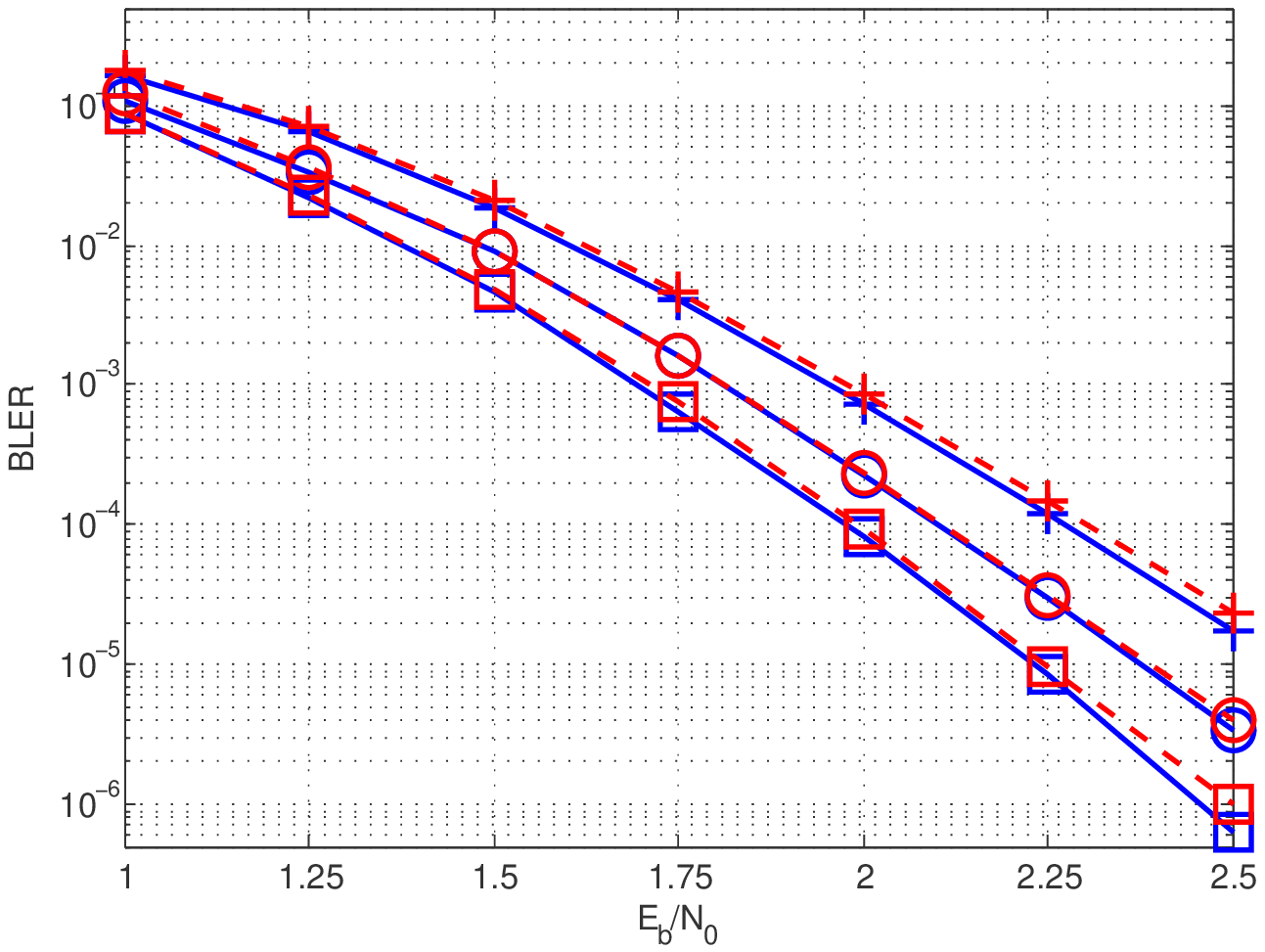}}

\subfloat{\includegraphics{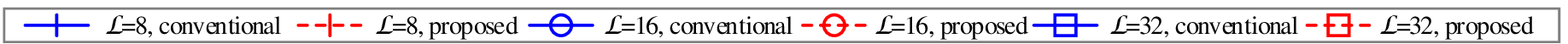}}

\caption{Error correction performance of the $(N,K,r)=(1024,512,24)$ polar
code decoded by SMB-DTS. }
\label{fig:err_perf}
\end{figure*}

\figref \ref{fig:err_perf} shows the BLERs of LSCDs with different
list sizes $\mathcal{L}$ (8, 16 and 32) and different numbers of
merged bits $M$ (2, 4 and 8) based on the floating-point simulation
results. The reliable set of the selective expansion scheme in the
SMB-DTS is obtained based on the method proposed in \cite{yzfan_hkust_jsac_2016_sedts}
with $\epsilon=30\%$ tolerable performance degradation compared with
that of the LSCD with $\mathcal{L}=32$ at $E_{b}/N_{0}=2.5$ dB.
For the DTS, we use the modified version, DTS-advance, presented in
\cite{yzfan_hkust_jsac_2016_sedts}, which is more suitable for hardware
implementation due to its lower computational complexity. The rejection
thresholds of the DTS-advance are $\pmc{6}{i}$, $\pmc{12}{i}$ and
$\pmc{25}{i}$ when $\mathcal{L}=8,\,16$ and 32, respectively. The
simulation results show that using SMB-DTS with different values of
$M$ does not have a significant impact on the error correction performance.
Based on the generated code sets, the numbers of SUBTs (SP1, SP2 and
rate-$\frac{1}{T}$ tuple) with different lengths under different
$M$ are obtained, as shown in \tabref \ref{tab:code_set}. It can
be seen even when $M=8$, out of the 128 eight-bit tuples, 90 tuples
are those for which all the eight bits can be decoded at the same
time, indicating a great latency-saving can be achieved, which will
be shown in the next sub-section. 
\begin{table}
\caption{Number of SUBTs with different lengths under different $M$}

\label{tab:code_set}\setlength{\tabcolsep}{3pt}

\begin{tabular}{c|c|c|c|c|c|c|c|c|c|c}
\hline 
\multirow{2}{*}{{\small{}$M$}} & \multirow{2}{*}{{\small{}1}} & \multicolumn{3}{c|}{{\small{}2}} & \multicolumn{3}{c|}{{\small{}4}} & \multicolumn{3}{c}{{\small{}8}}\tabularnewline
\cline{3-11} 
 &  & {\small{}SP1} & {\small{}SP2} & {\small{}$R$=$\frac{1}{T}$} & {\small{}SP1} & {\small{}SP2} & {\small{}$R$=$\frac{1}{T}$} & {\small{}SP1} & {\small{}SP2} & {\small{}$R$=$\frac{1}{T}$}\tabularnewline
\hline 
{\small{}2} & {\small{}34} & {\small{}64} & {\small{}377} & {\small{}54} & {\small{}-} & {\small{}-} & {\small{}-} & {\small{}-} & {\small{}-} & {\small{}-}\tabularnewline
\hline 
{\small{}4} & {\small{}34} & {\small{}32} & {\small{}3} & {\small{}30} & {\small{}32} & {\small{}159} & {\small{}24} & {\small{}-} & {\small{}-} & {\small{}-}\tabularnewline
\hline 
{\small{}8} & {\small{}34} & {\small{}32} & {\small{}3} & {\small{}30} & {\small{}17} & {\small{}5} & {\small{}13} & {\small{}15} & {\small{}64} & {\small{}11}\tabularnewline
\hline 
\end{tabular}
\end{table}
\begin{table}
\caption{Latency of the LSCDs with all the proposed schemes}

\label{tab:lat_lls_lscd}\setlength{\tabcolsep}{5pt}

\begin{tabular}{c|c|c|c|c|c|c}
\hline 
\multicolumn{2}{c|}{} & \multicolumn{3}{c|}{{\small{}This work}} & {\small{}\cite{abalatsoukas_epfl_tsp_2015_llrlscd}} & {\small{}\cite{yzfan_hkust_jsac_2016_sedts}}\tabularnewline
\hline 
\multicolumn{2}{c|}{{\small{}$M$}} & {\small{}2} & {\small{}4} & {\small{}8} & {\small{}-} & {\small{}-}\tabularnewline
\hline 
\hline 
\multicolumn{2}{c|}{{\small{}LM}} & {\small{}735} & {\small{}520} & {\small{}430} & {\small{}512} & {\small{}422}\tabularnewline
\hline 
\multirow{2}{*}{{\small{}SCD}} & {\small{}$<$stg. 3} & {\small{}384} & {\small{}128} & {\small{}-} & {\small{}1792} & {\small{}768}\tabularnewline
\cline{2-7} 
 & {\small{}$\geq$stg. 3} & {\small{}168} & {\small{}168} & {\small{}168} & {\small{}288} & {\small{}288}\tabularnewline
\hline 
\multicolumn{2}{c|}{{\small{}Total (w/o saving)}} & {\small{}1287} & {\small{}816} & {\small{}598} & {\small{}2592} & {\small{}1478}\tabularnewline
\hline 
\hline 
\multirow{2}{*}{{\small{}Saving}} & {\small{}$\mathcal{D}_{\text{fine}}$} & {\small{}256} & {\small{}128} & {\small{}64} & {\small{}-} & {\small{}-}\tabularnewline
\cline{2-7} 
 & {\small{}$\mathcal{D}_{\text{zero}}$} & {\small{}88} & {\small{}41} & {\small{}18} & {\small{}-} & {\small{}-}\tabularnewline
\hline 
\multicolumn{2}{c|}{{\small{}Total (with saving)}} & {\small{}943} & {\small{}647} & {\small{}516} & {\small{}2592} & {\small{}1478}\tabularnewline
\hline 
\end{tabular}
\end{table}

\subsection{Latency-Saving Achieved by the Proposed Schemes\label{subsec:lat_cal}}

\begin{table*}
\caption{Synthesis results of the proposed LSCD architectures for (1024,512,24)
polar codes}
\label{tab:impl}

{\small{}}%
\begin{tabular}{c||>{\centering}p{0.8cm}|>{\centering}p{0.8cm}|>{\centering}p{0.8cm}||>{\centering}p{0.8cm}|>{\centering}p{0.8cm}|>{\centering}p{0.8cm}||>{\centering}p{0.8cm}|>{\centering}p{0.8cm}|>{\centering}p{0.8cm}}
\hline 
{\small{}List size $\mathcal{L}$} & \multicolumn{3}{c||}{{\small{}32}} & \multicolumn{3}{c||}{{\small{}16}} & \multicolumn{3}{c}{{\small{}8}}\tabularnewline
\hline 
{\small{}Merged bits $M$} & \multicolumn{1}{c|}{{\small{}2}} & \multicolumn{1}{c|}{{\small{}4}} & \multicolumn{1}{c||}{{\small{}8}} & \multicolumn{1}{c|}{{\small{}2}} & \multicolumn{1}{c|}{{\small{}4}} & \multicolumn{1}{c||}{{\small{}8}} & \multicolumn{1}{c|}{{\small{}2}} & \multicolumn{1}{c|}{{\small{}4}} & \multicolumn{1}{c}{{\small{}8}}\tabularnewline
\hline 
{\small{}Clock freq. (MHz)} & {\small{}495} & {\small{}465} & {\small{}417} & {\small{}676} & {\small{}617} & {\small{}488} & {\small{}769} & {\small{}685} & {\small{}556}\tabularnewline
\hline 
\multirow{1}{*}{{\small{}Throughput (Mbps)}} & {\small{}537} & {\small{}736} & {\small{}827} & {\small{}734} & {\small{}977} & {\small{}968} & {\small{}835} & {\small{}1084} & {\small{}1103}\tabularnewline
\hline 
{\small{}Total area ($\text{mm}^{2}$)} & {\small{}18.26} & {\small{}18.76} & {\small{}19.58} & {\small{}7.85} & {\small{}8.01} & {\small{}8.07} & {\small{}4.38} & {\small{}4.43} & {\small{}4.54}\tabularnewline
\hline 
{\small{}Total area w/o PCMS ($\text{mm}^{2}$)} & {\small{}21.31} & {\small{}21.81} & {\small{}22.64} & {\small{}9.30} & {\small{}9.46} & {\small{}9.53} & {\small{}4.96} & {\small{}5.01} & {\small{}5.12}\tabularnewline
\hline 
\end{tabular}{\small \par}
\end{table*}
\begin{table}
\caption{Area breakdown of the LSCDs with $M=8$ (unit: $mm^{2}$)}
\label{tab:breakdown}

\begin{tabular}{c|c|>{\centering}p{1.2cm}|>{\centering}p{1.2cm}|>{\centering}p{1.2cm}}
\hline 
\multicolumn{2}{c|}{{\small{}List size $\mathcal{L}$}} & {\small{}32} & {\small{}16} & {\small{}8}\tabularnewline
\hline 
\multicolumn{2}{c|}{{\small{}Total area}} & {\small{}19.58} & {\small{}8.07} & {\small{}4.54}\tabularnewline
\hline 
\multicolumn{2}{c|}{{\small{}PMU}} & {\small{}0.64} & {\small{}0.31} & {\small{}0.21}\tabularnewline
\hline 
\multirow{3}{*}{{\small{}SCD cores}} & {\small{}LLR mem.} & {\small{}6.34} & {\small{}3.17} & {\small{}1.59}\tabularnewline
\cline{2-5} 
 & {\small{}PE array} & {\small{}1.83} & {\small{}0.95} & {\small{}0.82}\tabularnewline
\cline{2-5} 
 & {\small{}Crossbar} & {\small{}4.43} & {\small{}1.06} & {\small{}0.37}\tabularnewline
\hline 
\multicolumn{2}{c|}{{\small{}Other}} & {\small{}6.34} & {\small{}2.58} & {\small{}1.55}\tabularnewline
\hline 
\end{tabular}
\end{table}

The overall decoding latency of LSCD with different $M$ values for
the 1024-bit polar code are obtained based on \eqref{eq:lat_tot}
and are summarized in \tabref \ref{tab:lat_lls_lscd}. We assume
$P=64$ PEs are used in each SCD, which is the same as that used in
\cite{yzfan_hkust_jsac_2016_sedts,abalatsoukas_epfl_tsp_2015_llrlscd,jlin_lehigh_tvlsi_2016_highthpt,sahashemi_mcgill_tsp_2017_fastflexible,sahashemi_mcgill_tcasi_2016_ssclspc,sahashemi_mcgill_jetcas_2017_memeff}.
The parameters of SMB-DTS are the same as those presented in \secref
\ref{subsec:error_perf}. When more bits are merged, the latency of
LM using SMB-DTS and the overall latency are both decreased. The latency
of SCD using P-GLAH is also decreased as more stages in the LSCD are
calculated by the LM module instead of the SCD cores. The first information
bit is $u_{127}$ and $\mathcal{D}_{\text{zero}}$ is obtained accordingly. 

We also compare the latency with those of the state-of-the-art LSCDs
\cite{abalatsoukas_epfl_tsp_2015_llrlscd,yzfan_hkust_jsac_2016_sedts}.
For a fair comparison, the latency is re-calculated using the same
code setting mentioned in \secref \ref{subsec:error_perf}. When
$M=8$, the latency without the two further saving refinement schemes
is 598 cycles, which is 77\% and 60\% less than the latency of \cite{abalatsoukas_epfl_tsp_2015_llrlscd}
and \cite{yzfan_hkust_jsac_2016_sedts}, respectively. It seems that
most of the latency saving is achieved by P-GLAH used in the SCD part
while the latency of LM is almost not reduced. However, the proposed
SMB-DTS executes LM and SCD calculations below stage $m$ as a whole.
Thus for a fair comparison, the latency of SMB-DTS with $M=8$ should
be compared with the sum of the latency of LM and SCD calculations
below stage 3, which are 2304 and 1190 clock cycles in \cite{abalatsoukas_epfl_tsp_2015_llrlscd}
and \cite{yzfan_hkust_jsac_2016_sedts}, respectively. This means
the latency reduction achieved by SMB-DTS is 82\% and 64\% compared
with \cite{abalatsoukas_epfl_tsp_2015_llrlscd} and \cite{yzfan_hkust_jsac_2016_sedts},
respectively. For the remaining SCD calculations at stages equal to
or higher than stage 3, the latency saving achieved by P-GLAH is 42\%,
which is consistent with the theoretical analysis in \secref \ref{subsec:pglah_parallelism}.
When the two further saving refinement schemes are used, an even higher
latency reduction is achieved.

It can be seen that the more the merged bits are, the smaller the
latency of SMB-DTS is. However, a larger $M$ means that a more complex
PMU block has to be used, which incurs a larger area and longer critical
path. Moreover, the programmable PE used in P-GLAH and the de-pipelining
for the fine-tuning optimization also increase the logic delay. To
this end, a careful tradeoff should be made between a higher clock
frequency and fewer decoding cycles to achieve an optimal decoding
throughput according to the implementation results to be shown in
the next sub-section. 

\subsection{Implementation Results of the Proposed Architecture\label{subsec:impl_rslt}}

The proposed LSCD architecture for the polar code of $(N,K,r)=(1024,512,24)$
with the settings (the number of PEs used and the parameters of SMB-DTS)
presented in \secref \ref{subsec:error_perf} and \ref{subsec:lat_cal}
is implemented. For a fair comparison, the same quantization schemes
in \cite{yzfan_hkust_jsac_2016_sedts,abalatsoukas_epfl_tsp_2015_llrlscd,jlin_lehigh_tvlsi_2016_highthpt,sahashemi_mcgill_tsp_2017_fastflexible,sahashemi_mcgill_tcasi_2016_ssclspc,sahashemi_mcgill_jetcas_2017_memeff},
i.e., $Q_{\text{LLR}}=6$ and $Q_{\text{PM}}=8$, are used in our
implementation. The proposed LSCD is synthesized with a UMC 90 nm
CMOS process using Synopsys Design Compiler. The reported throughputs
are coded throughput. The reported area includes both cell and net
area.
\begin{table*}
{\small{}\caption{Synthesis results comparison with the state-of-the-art LSCD architectures
for 1024-bit polar codes}
\label{tab:comp}}{\small \par}

{\small{}}%
\begin{tabular}{>{\centering}p{4cm}||c|c|c|c|c|c|c|c|c}
\hline 
 & \multicolumn{3}{c|}{{\small{}This work ($M=8$)}} & {\small{}\cite{yzfan_hkust_jsac_2016_sedts}} & {\small{}$\,$\cite{sahashemi_mcgill_tsp_2017_fastflexible}$\dagger$} & {\small{}$\,$\cite{sahashemi_mcgill_tcasi_2016_ssclspc}$\dagger$} & {\small{}$\,$\cite{jlin_lehigh_tvlsi_2016_highthpt}$\diamond$} & {\small{}$\,$\cite{abalatsoukas_epfl_tsp_2015_llrlscd}$\diamond$} & {\small{}$\,$\cite{sahashemi_mcgill_jetcas_2017_memeff}$\dagger$}\tabularnewline
\hline 
{\small{}$K=|\mathcal{A}|$} & \multicolumn{3}{c|}{{\small{}512}} & {\small{}528} & {\small{}528$^{\triangle}$} & {\small{}528$^{\triangle}$} & {\small{}512} & {\small{}528$^{\triangle}$} & {\small{}544$^{\triangle}$}\tabularnewline
\hline 
{\small{}List size $\mathcal{L}$} & \multicolumn{1}{c|}{{\small{}32}} & {\small{}16} & {\small{}8} & {\small{}16} & {\small{}8} & {\small{}8} & {\small{}8} & {\small{}8} & {\small{}4}\tabularnewline
\hline 
{\small{}Clock freq. (MHz)} & {\small{}417} & {\small{}488} & {\small{}556} & {\small{}658} & {\small{}520} & {\small{}521} & {\small{}289} & {\small{}637} & {\small{}578}\tabularnewline
\hline 
{\small{}Throughput (Mbps)} & {\small{}827} & {\small{}968} & {\small{}1103} & {\small{}460} & {\small{}862} & {\small{}622} & {\small{}732} & {\small{}245} & {\small{}217}\tabularnewline
\hline 
{\small{}Total area ($\text{mm}^{2}$)} & {\small{}19.58} & {\small{}8.07} & {\small{}4.54} & {\small{}7.47} & {\small{}7.64} & {\small{}5.96} & {\small{}7.22} & {\small{}3.85} & {\small{}1.26}\tabularnewline
\hline 
{\small{}Area efficiency (Mbps/$\text{mm}^{2}$)} & {\small{}42.34} & {\small{}119.95} & {\small{}242.95} & {\small{}61.58} & {\small{}112.83} & {\small{}104.36} & {\small{}101.36} & {\small{}63.64} & {\small{}172.22}\tabularnewline
\hline 
\multicolumn{10}{l}{{\footnotesize{}$\dagger$ The synthesis results in \cite{sahashemi_mcgill_tsp_2017_fastflexible},
\cite{sahashemi_mcgill_tcasi_2016_ssclspc} and \cite{sahashemi_mcgill_jetcas_2017_memeff}
are based on TSMC 65 nm technology and are scaled to a 90 nm technology.}}\tabularnewline
\multicolumn{10}{l}{{\footnotesize{}$\diamond$ The synthesis results in \cite{abalatsoukas_epfl_tsp_2015_llrlscd}
and \cite{jlin_lehigh_tvlsi_2016_highthpt} are based on TSMC 90 nm
technology.}}\tabularnewline
\multicolumn{10}{l}{{\footnotesize{}$^{\triangle}$ The cardinalty $|\mathcal{A}|$ is
re-calculated according to the definition in this paper.}}\tabularnewline
\end{tabular}{\small \par}
\end{table*}

\tabref \ref{tab:impl} shows the synthesis results with different
list sizes and different numbers of merged bits. The maximum throughput
of LSCD with $\mathcal{L}=32$ is about 25\% lower than that of $\mathcal{L}=8$.
The throughputs of the proposed LSCD architectures with $M=4$ and
8 are similar and are much higher than that of $M=2$ although the
clock frequency is slightly reduced when a larger $M$ is used. The
critical paths of the implementations with $M=4$ and 8 both lie at
the datapath shown in \figref \ref{fig:arch_intf}(b), which is mainly
due to the logic delay of the PMU block. The critical paths of the
implementations with $M=2$ lies at the DTS block as the complexity
of the PMU block is small.

The area of the proposed architecture is greatly increased when a
large list size is used. This is mainly due to the crossbar in the
SCD cores. According to the area breakdown shown in \tabref \ref{tab:breakdown},
the area of the crossbar for $\mathcal{L}=32$ is four times that
of the crossbar for $\mathcal{L}=16$. The complex interconnection
of the crossbar also leads to a large area for routing. The complexity
of the PMU block is greatly increased with $M$ according to the discussion
in \secref \ref{subsec:pmu_str}. However, its area is only slightly
increased for a large $M$ because the PMU block contributes less
than 5\% of the total area. In contrast, the SCD cores contribute
about 65\% of the total area where the LLR memory is the main contributor.
The area-saving achieved by PCMS is 12.8\%, 18.1\% and 15.6\% for
LSCD with $\mathcal{L}=8$, 16 and 32, respectively, which are larger
than the 8\% saving for LSCD with $\mathcal{L}=4$ reported in \cite{crxiong_lehigh_tsp_2016_symbol}.
This shows that PCMS saves the area more effectively for a large list
size.

\tabref \ref{tab:comp} compares the performance of our architecture
with some state-of-the-art LSCD architectures \cite{yzfan_hkust_jsac_2016_sedts,abalatsoukas_epfl_tsp_2015_llrlscd,jlin_lehigh_tvlsi_2016_highthpt,sahashemi_mcgill_tsp_2017_fastflexible,sahashemi_mcgill_tcasi_2016_ssclspc,sahashemi_mcgill_jetcas_2017_memeff}.
It can be seen that the proposed architecture can support LSCDs with
larger list sizes and both the decoding throughput and the area efficiency
are higher than those of the state-of-the-art LSCDs with the same
list sizes. These higher throughputs are achieved under a similar
clock frequency, which means our architecture reduces the decoding
latency significantly. Comparing with \cite{yzfan_hkust_jsac_2016_sedts},
the proposed LSCDs with $\mathcal{L}=16$ have a similar area. The
area of the PE array and PMU is much larger than their counterparts
in \cite{yzfan_hkust_jsac_2016_sedts}, which are 0.53 $\text{mm}{}^{2}$
for the PE arrays and less than 0.1 $\text{mm}{}^{2}$ for the whole
LM module for an LSCD with $\mathcal{L}=16$, indicating the high
throughput is achieved at the cost of increased hardware complexity
of the programmable PE and PMU block. Fortunately, the total area
is not increased much as the area-saving brought by the PCMS offsets
the area overhead of the combinational logic. Comparing with \cite{abalatsoukas_epfl_tsp_2015_llrlscd},
although the area of our design is larger, a 3.5 times area efficiency
is achieved due to the shorter decoding latency, as stated in \secref
\ref{subsec:lat_cal}. Comparing with the fast-SSC-based architectures
\cite{sahashemi_mcgill_tsp_2017_fastflexible,sahashemi_mcgill_tcasi_2016_ssclspc},
fewer path expansions are required for SP1 and SP2 in SMB-DTS comparing
with the rate-1 codes, leading to a higher throughput and area efficiency.

\section{Conclusion\label{sec:conclusion}}

In this work, a high-throughput architecture for the LSCD with a large
list size is proposed. First, two kinds of low-latency decoding algorithms
are proposed. For the list management module, a multi-bit double thresholding
scheme is proposed so that the double thresholding scheme can work
with multi-bit decoding to reduce the latency. For the SCD cores,
a partial G-node look-ahead scheme is proposed by making a tradeoff
between the complexity and the latency. A high-performance VLSI architecture
is then developed based on the proposed algorithms. Experimental results
show that LSCDs with $\mathcal{L}=8$, 16 and 32 implemented by the
proposed architecture provide much higher throughputs than the state-of-the-art
architectures with a good BLER performance.


\begin{thebibliography}{10}
\providecommand{\url}[1]{#1}
\csname url@samestyle\endcsname
\providecommand{\newblock}{\relax}
\providecommand{\bibinfo}[2]{#2}
\providecommand{\BIBentrySTDinterwordspacing}{\spaceskip=0pt\relax}
\providecommand{\BIBentryALTinterwordstretchfactor}{4}
\providecommand{\BIBentryALTinterwordspacing}{\spaceskip=\fontdimen2\font plus
\BIBentryALTinterwordstretchfactor\fontdimen3\font minus
  \fontdimen4\font\relax}
\providecommand{\BIBforeignlanguage}[2]{{%
\expandafter\ifx\csname l@#1\endcsname\relax
\typeout{** WARNING: IEEEtran.bst: No hyphenation pattern has been}%
\typeout{** loaded for the language `#1'. Using the pattern for}%
\typeout{** the default language instead.}%
\else
\language=\csname l@#1\endcsname
\fi
#2}}
\providecommand{\BIBdecl}{\relax}
\BIBdecl

\bibitem{earikan_bilkent_tit_2009_polar}
E.~Ar{\i}kan, ``Channel polarization: A method for constructing
  capacity-achieving codes for symmetric binary-input memoryless channels,''
  \emph{{IEEE} Trans. Inf. Theory}, vol.~55, no.~7, pp. 3051--3073, June 2009.

\bibitem{ital_ucsd_tit_2013_construction}
I.~Tal and A.~Vardy, ``How to construct polar codes,'' \emph{{IEEE} Trans. Inf.
  Theory}, vol.~59, no.~10, pp. 6562--6582, Oct 2013.

\bibitem{amishra_epfl_asscc_2012_asic}
A.~Mishra, A.~J. Raymond, L.~G. Amaru, G.~Sarkis, C.~Leroux, P.~Meinerzhagen,
  A.~Burg, and W.~J. Gross, ``A successive cancellation decoder {ASIC} for a
  1024-bit polar code in 180 nm {CMOS},'' in \emph{{IEEE} Asian Solid-State
  Circuits Conf. {(A-SSCC)}}, 2012, pp. 205--208.

\bibitem{cleroux_mcgill_tsp_2013_semiparallel}
C.~Leroux, A.~J. Raymond, G.~Sarkis, and W.~J. Gross, ``A semi-parallel
  successive-cancellation decoder for polar codes,'' \emph{{IEEE} Trans. Signal
  Process.}, vol.~61, no.~2, pp. 289--299, Jan 2013.

\bibitem{czhang_umn_tcasii_2014_simplified}
C.~Zhang and K.~K. Parhi, ``Latency analysis and architecture design of
  simplified {SC} polar decoders,'' \emph{{IEEE} Trans. Circuits Syst. {II}},
  vol.~61, no.~2, pp. 115--119, Feb 2014.

\bibitem{yzfan_hkust_tsp_2014_effps}
Y.~Fan and C.-Y. Tsui, ``An efficient partial-sum network architecture for
  semi-parallel polar codes decoder implementation,'' \emph{{IEEE} Trans.
  Signal Process.}, vol.~62, no.~12, pp. 3165--3179, Jun 2014.

\bibitem{czhang_umn_icc_2012_lookahead}
C.~Zhang, B.~Yuan, and K.~K. Parhi, ``Reduced-latency {SC} polar decoder
  architectures,'' in \emph{{IEEE} Int. Conf. Commun. {(ICC)}}, 2012, pp.
  3471--3475.

\bibitem{czhang_umn_tsp_2013_overlap}
C.~Zhang and K.~K. Parhi, ``Low-latency sequential and overlapped architectures
  for successive cancellation polar decoder,'' \emph{{IEEE} Trans. Signal
  Process.}, vol.~61, no.~10, pp. 2429--2441, May 2013.

\bibitem{byuan_umn_tcasi_2014_sc2bd}
B.~Yuan and K.~K. Parhi, ``Low-latency successive-cancellation polar decoder
  architectures using 2-bit decoding,'' \emph{{IEEE} Trans. Circuits Syst.
  {I}}, vol.~61, no.~4, pp. 1241--1254, Apr 2014.

\bibitem{gsarkis_mcgill_jsac_2014_fast}
G.~Sarkis, P.~Giard, A.~Vardy, C.~Thibeault, and W.~J. Gross, ``Fast polar
  decoders: Algorithm and implementation,'' \emph{{IEEE} J. Sel. Areas
  Commun.}, vol.~32, no.~5, pp. 946--957, May 2014.

\bibitem{apamuk_icwcs_bilkent_2011_fpga}
A.~Pamuk, ``An {FPGA} implementation architecture for decoding of polar
  codes,'' in \emph{{IEEE} Int. Symp. Wireless Commun. Syst. {(ISWCS)}}, 2011,
  pp. 437--441.

\bibitem{byuan_umn_tsp_2014_earlystop}
B.~Yuan and K.~K. Parhi, ``Early stopping criteria for energy-efficient
  low-latency belief-propagation polar code decoders,'' \emph{{IEEE} Trans.
  Signal Process.}, vol.~62, no.~24, pp. 6496--6506, Dec 2014.

\bibitem{smabbas_hkust_tvlsi_2016_bpd}
S.~M. Abbas, Y.~Fan, J.~Chen, and C.~Tsui, ``High-throughput and
  energy-efficient belief propagation polar code decoder,'' \emph{{IEEE} Trans.
  {VLSI} Syst.}, vol.~PP, no.~99, pp. 1--14, Early Access.

\bibitem{rggallager_mit_book_1963_ldpc}
R.~G. Gallager, \emph{Low-Density Parity-Check Codes}.\hskip 1em plus 0.5em
  minus 0.4em\relax Cambridge, MA, USA: MIT Press, 1963.

\bibitem{cberrou_bretagne_icc_1993_turbo}
C.~Berrou, A.~Glavieux, and P.~Thitimajshima, ``Near shannon limit
  error-correcting coding and decoding: Turbo-codes. 1,'' in \emph{{IEEE} Int.
  Conf. Commun. {(ICC)}}, May 1993, pp. 1064--1070.

\bibitem{kchen_bupt_iet_2012_lscd}
K.~Chen, K.~Niu, and J.~R. Lin, ``List successive cancellation decoding of
  polar codes,'' \emph{IET Electron. Lett.}, vol.~48, no.~9, pp. 500--501, Apr
  2012.

\bibitem{ital_ucsd_tit_2015_list}
I.~Tal and A.~Vardy, ``List decoding of polar codes,'' \emph{{IEEE} Trans. Inf.
  Theory}, vol.~61, no.~5, pp. 2213--2226, May 2015.

\bibitem{bli_huawei_cl_2012_crc}
B.~Li, H.~Shen, and D.~Tse, ``An adaptive successive cancellation list decoder
  for polar codes with cyclic redundancy check,'' \emph{{IEEE} Commun. Lett.},
  vol.~16, no.~12, pp. 2044--2047, Dec 2012.

\bibitem{kniu_bupt_cl_2012_crc}
K.~Niu and K.~Chen, ``{CRC}-aided decoding of polar codes,'' \emph{{IEEE}
  Commun. Lett.}, vol.~16, no.~10, pp. 1668--1671, Oct 2012.

\bibitem{kniu_bupt_icc_2013_beyondturbo}
K.~Niu, K.~Chen, and J.~R. Lin, ``Beyond turbo codes: Rate-compatible punctured
  polar codes,'' in \emph{Proc. IEEE Int. Conf. Commun.(ICC)}, 2013, pp.
  3423--3427.

\bibitem{3gpp_3gpp_ran087_2016_5g}
3rd Generation Partnership~Project, ``{Draft Report of 3GPP TSG RAN WG1 \#87
  v0.1.0},''
  \url{http://www.3gpp.org/ftp/tsg_ran/WG1_RL1/TSGR1_87/Report/Draft_Minutes_report_RAN1#87_v011.zip},
  p. 129, 2016, [Online; accessed 11-Jan-2017].

\bibitem{abalatsoukas_epfl_tcasii_2014_archlscd}
A.~Balatsoukas-Stimming, A.~J. Raymond, W.~J. Gross, and A.~Burg, ``Hardware
  architecture for list successive cancellation decoding of polar codes,''
  \emph{{IEEE} Trans. Circuits Syst. {II}}, vol.~61, no.~8, pp. 609--613, May
  2014.

\bibitem{byuan_umn_asilomar_2014_llrlscd}
B.~Yuan and K.~K. Parhi, ``Successive cancellation list polar decoder using
  log-likelihood ratios,'' in \emph{{IEEE} Asilomar Conf. Signals, Syst., and
  Computers {(ACSSC)}}, 2014, pp. 548--552.

\bibitem{abalatsoukas_epfl_icassp_2014_llrlscd}
A.~Balatsoukas-Stimming, M.~Bastani~Parizi, and A.~Burg, ``{LLR}-based
  successive cancellation list decoding of polar codes,'' in \emph{{IEEE} Int.
  Conf. Acoust., Speech, Signal Process. {(ICASSP)}}, 2014, pp. 3903--3907.

\bibitem{abalatsoukas_epfl_tsp_2015_llrlscd}
------, ``{LLR}-based successive cancellation list decoding of polar codes,''
  \emph{{IEEE} Trans. Signal Process.}, vol.~63, no.~19, pp. 5165--5179, Oct
  2015.

\bibitem{pgiard_epfl_jetcas_2017_polarbear}
P.~Giard, A.~Balatsoukas-Stimming, T.~Müller, A.~Bonetti, C.~Thibeault, W.~J.
  Gross, P.~Flatresse, and A.~Burg, ``{POLARBEAR}: A 28-nm {FD-SOI} {ASIC} for
  decoding of polar codes,'' \emph{{IEEE} Trans. Emerg. Sel. Topics Circuits
  Syst.}, vol.~PP, no.~99, pp. 1--14, 2017.

\bibitem{byuan_umn_tvlsi_2015_sclmbd}
B.~Yuan and K.~K. Parhi, ``Low-latency successive-cancellation list decoders
  for polar codes with multibit decision,'' \emph{{IEEE} Trans. {VLSI} Syst.},
  vol.~23, no.~10, pp. 2268--2280, Oct 2015.

\bibitem{byuan_umn_tcasii_2017_sclmbd}
------, ``{LLR}-based successive-cancellation list decoder for polar codes with
  multi-bit decision,'' \emph{{IEEE} Trans. Circuits Syst. {II}}, vol.~64,
  no.~1, pp. 21--25, Jan. 2017.

\bibitem{crxiong_lehigh_sips_2014_symbol}
C.~Xiong, J.~Lin, and Z.~Yan, ``Symbol-based successive cancellation list
  decoder for polar codes,'' in \emph{{IEEE} Workshop Signal Process. Syst.
  {(SiPS)}}, 2014, pp. 1--6.

\bibitem{crxiong_lehigh_tsp_2016_symbol}
------, ``Symbol-decision successive cancellation list decoder for polar
  codes,'' \emph{{IEEE} Trans. Signal Process.}, vol.~64, no.~3, pp. 675--687,
  Feb 2016.

\bibitem{sahashemi_mcgill_isit_2016_sscl}
S.~A. Hashemi, C.~Condo, and W.~J. Gross, ``Simplified successive-cancellation
  list decoding of polar codes,'' in \emph{{IEEE} Int. Symp. Inf. Theory
  {(ISIT)}}, 2016, pp. 815--819.

\bibitem{sahashemi_mcgill_tcasi_2016_ssclspc}
------, ``A fast polar code list decoder architecture based on sphere
  decoding,'' \emph{{IEEE} Trans. Circuits Syst. {I}}, vol.~63, no.~12, pp.
  2368--2380, Dec 2016.

\bibitem{sahashemi_mcgill_wcncw_2017_fastsscl}
------, ``Fast simplified successive-cancellation list decoding of polar
  codes,'' in \emph{{IEEE} Wireless Commun. and Networking Conf. Workshops
  {(WCNCW)}}, 2017, pp. 1--6.

\bibitem{sahashemi_mcgill_tsp_2017_fastflexible}
------, ``Fast and flexible successive-cancellation list decoders for polar
  codes,'' \emph{{IEEE} Trans. Signal Process.}, vol.~65, no.~21, pp. 5756 --
  5769, Nov 2017.

\bibitem{jlin_lehigh_sips_2014_rdclat}
J.~Lin, C.~Xiong, and Z.~Yan, ``A reduced latency list decoding algorithm for
  polar codes,'' in \emph{{IEEE} Workshop Signal Process. Syst. {(SiPS)}},
  2014, pp. 1--6.

\bibitem{jlin_lehigh_tvlsi_2016_highthpt}
------, ``A high throughput list decoder architecture for polar codes,''
  \emph{{IEEE} Trans. {VLSI} Syst.}, vol.~24, no.~6, pp. 2378--2391, June 2016.

\bibitem{gsarkis_mcgill_jsac_2016_sclfast}
G.~Sarkis, P.~Giard, A.~Vardy, C.~Thibeault, and W.~J. Gross, ``Fast list
  decoders for polar codes,'' \emph{{IEEE} J. Sel. Areas Commun.}, vol.~34,
  no.~2, pp. 318--328, Feb. 2016.

\bibitem{abalatsoukas_epfl_iscas_2015_sorting}
A.~Balatsoukas-Stimming, M.~Bastani~Parizi, and A.~Burg, ``On metric sorting
  for successive cancellation list decoding of polar codes,'' in \emph{{IEEE}
  Int. Symp. Circ. and Syst. {(ISCAS)}}, 2015, pp. 1993--1996.

\bibitem{yzfan_hkust_icassp_2015_dts}
Y.~Fan, J.~Chen, C.~Xia, C.-y. Tsui, J.~Jin, H.~Shen, and B.~Li, ``Low-latency
  list decoding of polar codes with double thresholding,'' in \emph{{IEEE} Int.
  Conf. Acoust., Speech, Signal Process. {(ICASSP)}}, 2015, pp. 1042--1046.

\bibitem{yzfan_hkust_jsac_2016_sedts}
Y.~Fan, C.~Xia, J.~Chen, C.~Tsui, J.~Jin, H.~Shen, and B.~Li, ``A low-latency
  list successive-cancellation decoding implementation for polar codes,''
  \emph{{IEEE} J. Sel. Areas Commun.}, vol.~34, no.~2, pp. 303--317, Feb. 2016.

\bibitem{cxia_hkust_fpl_2017_fpgalarge}
C.~Xia, Y.~Fan, J.~Chen, C.~Tsui, C.~Zeng, J.~Jin, and B.~Li, ``An
  implementation of list successive cancellation decoder with large list size
  for polar codes,'' in \emph{Int. Conf. Field Programmable Logic and Appl.
  (FPL)}, 2017, pp. 1--4.

\bibitem{zyzhang_zju_jsac_2016_splitreduce}
Z.~Zhang, L.~Zhang, X.~Wang, C.~Zhong, and H.~V. Poor, ``A split-reduced
  successive cancellation list decoder for polar codes,'' \emph{{IEEE} J. Sel.
  Areas Commun.}, vol.~34, no.~2, pp. 292--302, Feb. 2016.

\bibitem{tbchen_tamu_iscas_2016_scloverlap}
T.~Chen, J.~Xu, and G.~Choi, ``Overlapped list successive cancellation approach
  for hardware efficient polar code decoder,'' in \emph{{IEEE} Int. Symp. Circ.
  and Syst. {(ISCAS)}}, 2016, pp. 2463--2466.

\bibitem{mhjang_snu_isita_2006_ldpcpfm}
M.-H. Jang, B.~Shiny, W.-M. Parky, J.-S. Noy, and I.-S. Jeon, ``On the decoding
  of {LDPC} codes in {IEEE} 802.16e standards for improving the convergence
  speed,'' in \emph{{IEEE} Int. Symp. Inf. Theory and Appl.{(ISITA)}}, 2006,
  pp. 624--628, [Online; accessed 11-Jul-2017].

\bibitem{bli_huawei_istc_2014_hybrid}
B.~Li, H.~Shen, D.~Tse, and W.~Tong, ``Low-latency polar codes via hybrid
  decoding,'' in \emph{{IEEE} Int. Symp. Turbo Codes and Iter. Inf. Process.
  {(ISTC)}}, 2014, pp. 223--227.

\bibitem{sahashemi_mcgill_jetcas_2017_memeff}
S.~A. Hashemi, C.~Condo, F.~Ercan, and W.~J. Gross, ``Memory-efficient polar
  decoders,'' \emph{{IEEE} Trans. Emerg. Sel. Topics Circuits Syst.}, vol.~PP,
  no.~99, pp. 1--12, 2017.

\end{thebibliography}

\end{document}